\documentclass[journal,twoside]{IEEEtran}

\usepackage{amsmath,amssymb,amsfonts}
\usepackage[ruled,vlined,linesnumbered]{algorithm2e}
\usepackage{algorithmic}
\usepackage{graphicx}
\usepackage{textcomp}
\usepackage{url}
\usepackage{float}
\usepackage{mathrsfs}
\usepackage{hyperref}
\usepackage{pdfpages}
\newcommand\blfootnote[1]{%
  \begingroup
  \renewcommand\thefootnote{}\footnote{#1}%
  \addtocounter{footnote}{-1}%
  \endgroup
}

\begin{document}

\title{User Training with Error Augmentation for sEMG-based Gesture Classification}
\author{
Yunus Bicer$^{\dagger}$,
Niklas Smedemark-Margulies$^{\dagger}$,
Basak Celik,
Elifnur Sunger,
Ryan Orendorff,
Stephanie Naufel,
Tales Imbiriba,
Deniz Erdo\u{g}mu\c{s},
Eugene Tunik, and
Mathew Yarossi
\\
\footnotesize$^{\dagger}$Equal Contribution
\thanks{
Manuscript received 14 September 2023; revised 23 January 2024;
accepted 23 February 2024. This work was supported by Meta Reality
Labs Research. (Yunus Bicer and Niklas Smedemark-Margulies contributed equally to this work.) (Corresponding author: Mathew Yarossi.)
This work involved human subjects or animals in its research.
Approval of all ethical and experimental procedures and protocols was
granted by Northeastern University Institutional Review Board under IRB
No. 15-10-22.
Yunus Bicer, Basak Celik, Elifnur Sunger, and Deniz Erdo\u{g}mu\c{s} are
with the Department of Electrical and Computer Engineering, Northeastern University, Boston, MA 02115 USA.
Niklas Smedemark-Margulies is with the Khoury College of Computer
Sciences, Northeastern University, Boston, MA, USA.
Stephanie Naufel and Ryan Orendorff are with Meta Reality Labs Research, Menlo Park, CA 94025 USA.
Tales Imbiriba is with the Department of Electrical and Computer Engineering and the Institute for Experiential AI, Northeastern University,
Boston, MA 02115 USA.
Eugene Tunik is with the Department of Physical Therapy, Movement,
and Rehabilitation Sciences, and the Institute for Experiential AI, Northeastern University, Boston, MA 02115 USA.
Mathew Yarossi is with the Department of Electrical and Computer
Engineering, and the Department of Physical Therapy, Movement, and
Rehabilitation Sciences, Northeastern University, Boston, MA 02115 USA (e-mail: m.yarrosi@northeastern.edu). 
}
}

\IEEEoverridecommandlockouts
\IEEEpubid{\makebox[\columnwidth]{\copyright2024 IEEE. DOI: \href{https://doi.org/10.1109/TNSRE.2024.3372512}{10.1109/TNSRE.2024.3372512}
\hfill} \hspace{\columnsep}\makebox[\columnwidth]{ }}
\maketitle
\IEEEpubidadjcol

\begin{abstract}
We designed and tested a system for real-time control of a user interface by extracting surface electromyographic (sEMG) activity from eight electrodes in a wristband configuration. sEMG data were streamed into a machine-learning algorithm that classified hand gestures in real-time. After an initial model calibration, participants were presented with one of three types of feedback during a human-learning stage: veridical feedback, in which predicted probabilities from the gesture classification algorithm were displayed without alteration; modified feedback, in which we applied a hidden augmentation of error to these probabilities; and no feedback. User performance was then evaluated in a series of minigames, in which subjects were required to use eight gestures to manipulate their game avatar to complete a task. Experimental results indicated that relative to the baseline, the modified feedback condition led to significantly improved accuracy. Class separation also improved, though this trend was not significant.
These findings suggest that real-time feedback in a gamified user interface with manipulation of feedback may enable intuitive, rapid, and accurate task acquisition for sEMG-based gesture recognition applications.\blfootnote{Code and data are available on-line at: \url{https://github.com/neu-spiral/emg-feedback-user-training}}
\end{abstract}

\begin{IEEEkeywords}
Myoelectric control,
Gesture recognition,
Human-computer interaction,
Error augmentation,
Co-adaptation,
Surface Electromyography (sEMG),
\end{IEEEkeywords}

\section{Introduction} \label{sec:Introduction}
Surface electromyography (sEMG) provides a convenient sensor modality for human-computer interaction (HCI) applications \cite{yang2021dynamic}. In the past two decades, research efforts have sought to translate the electrical activity associated with muscle contraction into control commands for general use computing, prosthetic control, and motor rehabilitation \cite{xiong2021deep,qi2019intelligent}. As the demand for more intuitive and responsive interfaces has grown, the focus on sEMG-based gesture recognition has intensified. 

Traditional approaches to sEMG-based gesture recognition assumed stationarity of the mapping between muscle activation and gestures, and did not consider the user's ability to adapt their behavior based on feedback about gesture classification performance. The emergence of co-adaptive learning algorithms in the past decade represented a marked shift, acknowledging both human and machine learning as parts of an integrated system \cite{de2021framework, hahne2015concurrent, couraud2018model, yeung2019directional, yeung2022co, nawfel2022influence}. 
One key finding from these approaches is that when the human receives continuous feedback about the mapping of muscle activation to gesture, they can increase classification performance through behavioral adaptations \cite{krasoulis2019effect, de2020guiding}. These adaptations can result in increased class separability \cite{bunderson2012quantification} and increased movement repeatability \cite{powell2013user}. However, the relationship between feature space adaptations and classifier performance is complex. Increased real-time classifier performance has also been found even in the absence of EMG feature space changes in relative class distributions \cite{franzke2020exploring}.  Despite the complex relationship between feature space class distributions and classifier performance, the influence of human learning on myoelectric gesture classification remains a compelling target of investigation.  

Human learning about myoelectric gesture classification can be considered as a form of motor skill learning. In the literature on motor learning, the canonical view is that humans use a combination of intrinsic feedback (sensory information) and augmented feedback (information that is not readily accessible through intrinsic feedback) \cite{magill2010motor}. 
Augmented feedback can be further categorized as providing `knowledge of performance' (information about specific movements and muscle activations), or `knowledge of results' (information about outcomes) \cite{lauber2014improving, sharma2016effectiveness}. 
In the present study, we focus on myoelectric control, where providing knowledge of results corresponds to providing output from a classifier, while knowledge of performance corresponds to descriptions of the features extracted from the sEMG. The ability to shape human behavior in traditional motor skill learning settings through the use of augmented feedback is well established. Strategies such as error augmentation \cite{wei2005visual, todorov1997augmented, hasson2016neuromotor} and reward manipulation \cite{haith2012evidence, huber2016persistence} have been shown to affect the rate and retention of learning as well as behavioral variability. Yet, to our knowledge, the use of error-augmented feedback has not been tested for co-adaptation approaches to sEMG-based gesture recognition. 

In this study, we conducted an experiment to test whether modified feedback about class posterior probabilities affects performance in a myoelectric control task.
We provided subjects with a form of error-augmented knowledge of results; by altering class probabilities, we diminished the differences between classes, making it harder for the target gesture class to exceed a predefined decision threshold. In particular, we softened probabilities towards a uniform distribution. This form of feedback manipulation is closely related to previous uses of error augmentation, also referred to as error amplification \cite{parmar2022sensory,moinuddin2021role,israely2016error}.
As mentioned, this form of feedback has been shown to hasten learning and improve the quality of self-evaluation \cite{wei2005visual,patton2013visuomotor} and increase retention of learned skills \cite{losey2016improving,parmar2022sensory}.
We therefore hypothesized that error amplification by softening probabilities would increase subsequent gesture classification performance by enhancing human skill learning. 
The knowledge gained from this investigation has broad potential applications for use in myoelectric prosthetics, assistive devices, and human-computer interfaces where users perform only a brief 4-minute calibration, and human learning may be critical to the success of model performance. 

\section{Experimental Design} \label{sec:Experimental_Design}
All protocols were approved by the Northeastern University Institutional Review Board (IRB number 15-10-22) in conformance with the declaration of Helsinki. 

\subsection{Subjects} \label{sec:Subjects}
Forty-four right-handed subjects (21 male / 23 female, mean age $\pm$ 1 standard deviation: $20.9 \pm 4.3$ years) participated after providing IRB-approved written informed consent. Subjects were free of orthopedic or neurological diseases that could interfere with the task and had normal or corrected-to-normal vision. 

\subsection{Experimental Setup} \label{sec:Experimental_Setup}

\begin{figure}
    \centering
    \includegraphics[width=0.8\linewidth]{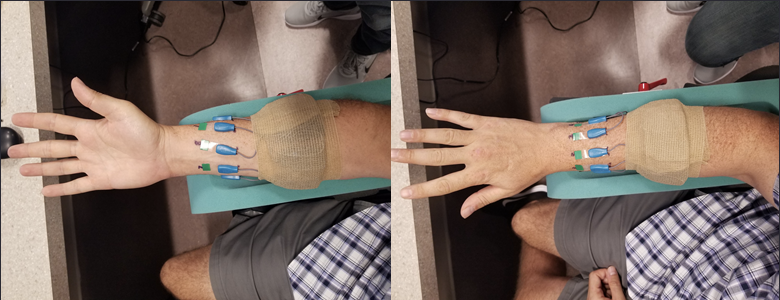}
    \caption{Electrode Placement. sEMG data is collected using $8$ Delsys Trigno sEMG sensors uniformly spaced around the right forearm.}
    \label{fig:data-collection}
\end{figure}

Subjects viewed a computer display while seated at a table with their right arm positioned comfortably in an armrest trough. 
Surface electromyography (sEMG) (Trigno, Delsys Inc., sampling frequency: $1926$ Hz) was collected from the muscles of the right forearm. 
Eight sEMG electrodes were placed at equidistant positions around the circumference of the forearm, at a four finger-width distance from the ulnar styloid (the subject's left hand was wrapped around the right forearm at the ulnar styloid to determine the sEMG placement). 
The first electrode was placed mid-line on the dorsal aspect of the forearm, and the other electrodes were then equally spaced (see Figure~\ref{fig:data-collection}). 

\subsection{Data Acquisition} \label{sec:Data_Collection_Procedure}

\subsubsection{Subject Group Assignment} 
Subjects were randomly assigned to one of three groups and performed a series of tasks as described below. 
Subjects who were unable to complete all tasks were excluded from further analysis.
Each subject group was assigned a different feedback condition: no feedback (``Control", N=$14$), veridical feedback (``Veridical", N=$14$), or modified feedback (``Modified", N=$16$) (see Section~\ref{sec:Block_Three_Live_Feedback} for details).
Subject group assignments were randomized before enrollment. In order to control for the possible confounding effect of biological variation in baseline performance across groups, we adopted a within-subject normalization strategy (see Section~\ref{sec:baselines}). 

\subsubsection{Gesture Timing} \label{sec:Gesture_Timing}
\begin{figure}[htb]
    \centering
    \includegraphics[width=1.0\linewidth,clip,trim=18 22 14 16]{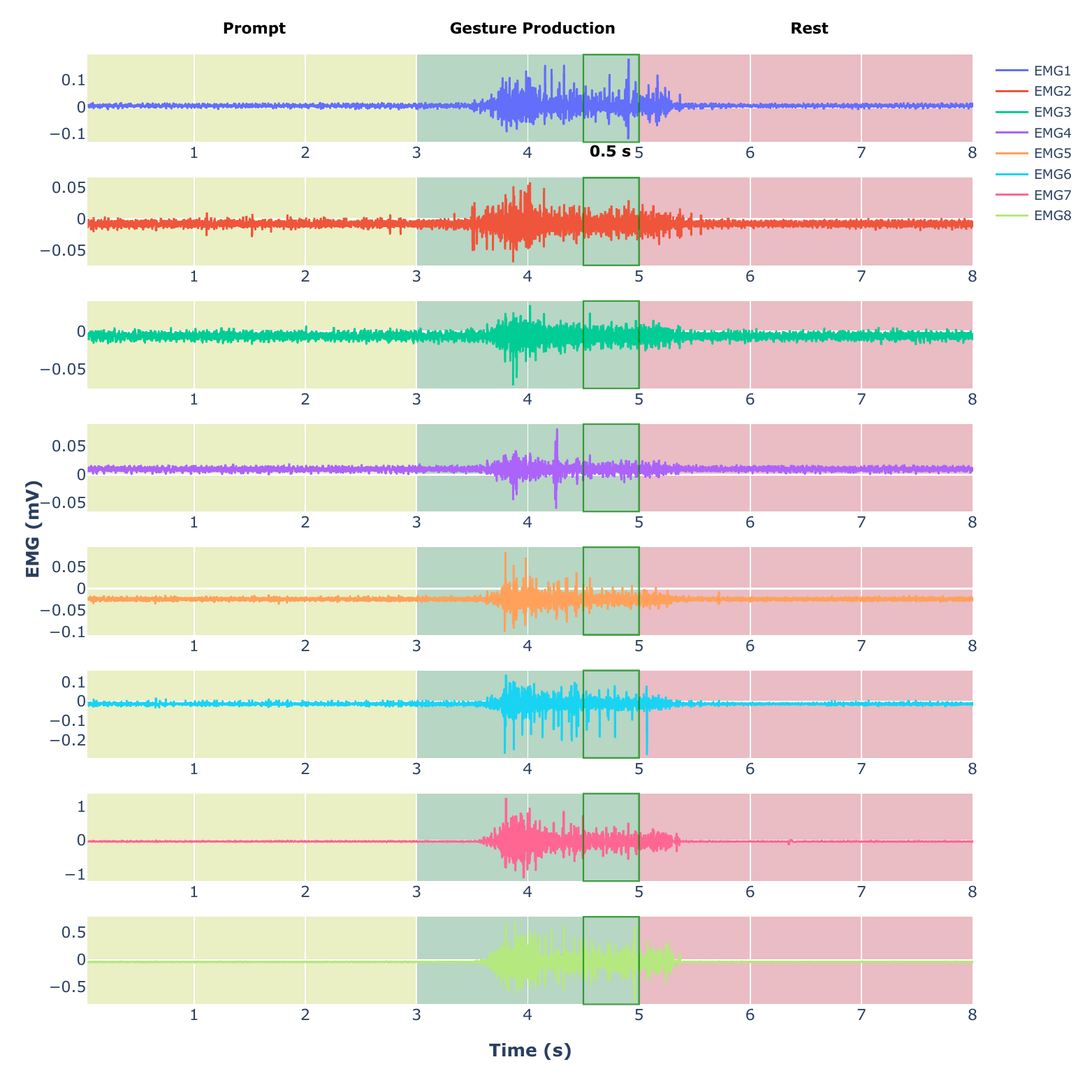}
    \caption{Gesture Trial Timing. In the yellow `prompting' epoch, the subject sees an instruction. In the green `gesture production' epoch, the subject performs the gesture. In the red `recovery' epoch, the subject returns to the rest position. Features for classification are extracted from the last $500$ ms of gesture production to help ensure that steady-state features are collected.}
    \label{fig:YGR-window}
\end{figure}

Subjects performed a series of tasks composed of one or more gesture trials to move an avatar dice (see details of user interface below). 
Prior to the start of a trial, the subject's forearm and wrist rested in a pronated position on the trough with the wrist neutral. In each trial, subjects were required to rest or to produce one of eight active gestures (label and action provided in brackets): index-thumb pinch [``Pinch", decrease number on avatar dice], index-thumb key press [``Thumb", increase the number on avatar dice], closed fist [``Fist", decrease size of avatar dice], full finger extension [``Open", increase size of avatar dice], wrist extension [``Up", move up], wrist flexion [``Down", move down], wrist radial deviation [``Left", move left], wrist ulnar deviation [``Right", move right]. 
Each trial began with a `prompting' epoch ($3$ sec) cued by a yellow bounding box the participant's display and a picture of the instructed gesture (Calibration and Instructed blocks only, see below), a `gesture production' epoch ($2$ sec) cued by a green bounding box, and a `recovery' epoch ($3$ sec) cued by a red bounding box. 
The final $500$ milliseconds of the gesture production epoch were used for feature extraction and classification.
Figure~\ref{fig:YGR-window} shows the timing of an example gesture trial.
This trial timing structure was chosen empirically to give enough time for subjects to prepare for each upcoming trial while keeping the total experiment duration short. Gesture trial timing was kept consistent to ensure that subject reaction times were not a source of variation in performance.

Each experimental session was divided into four blocks. 
Blocks one, two, and four used the trial timing described above. By contrast, in block three (in which some subjects received model feedback) the gesture production epoch lasted $30$ seconds for each gesture. 
During this time period, continuous feedback was provided by applying a classifier model on a sliding window of data, with a step size of $13.5$ milliseconds (based on the frequency of data packets delivered by our sEMG sensors).

\subsubsection{Block One: Calibration} \label{sec:Block_One_Calibration}
Subjects from all groups were instructed to perform five consecutive repetitions of each active gesture and eight repetitions of a rest gesture in which they were asked to relax the hand. This consecutive structure was chosen to help keep the task simple while the participant initially learned the set of available gestures. A classification model was trained on this small dataset before continuing to the next experimental block.

\subsubsection{Block Two: Instructed Games} \label{sec:Block_Two_Instructed_Games}
\begin{figure}[htb]
    \centering
    \includegraphics[width=0.8\linewidth,clip,trim=100 600 700 300]{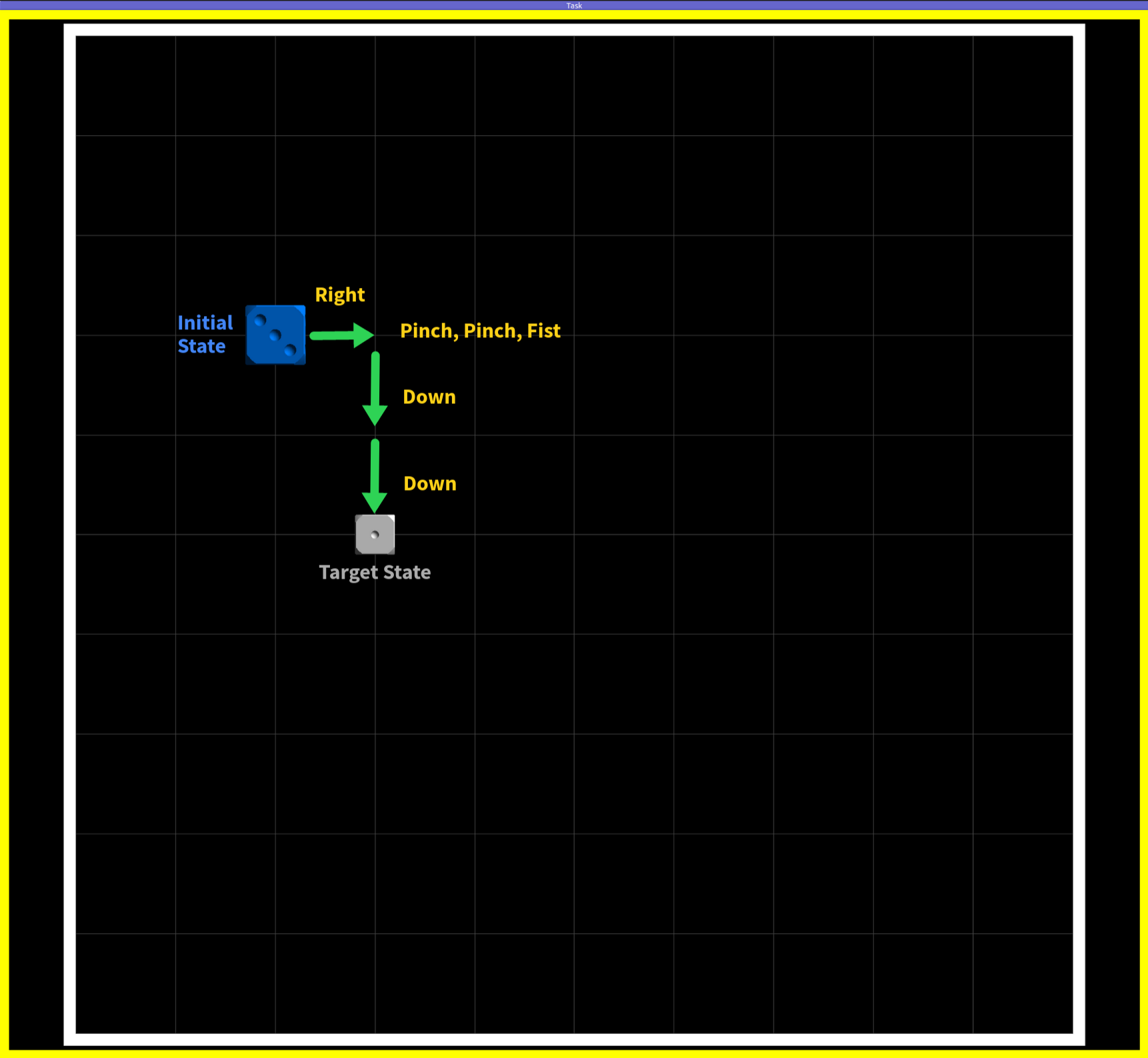}
    \caption{Example mini game. The blue player avatar must be moved to match the gray target avatar. The minimal path includes moving right, down twice, decreasing the die number (using a pinch gesture), and reducing size (using a fist gesture). }
    \label{fig:mini-game}
\end{figure}

Subjects from all groups engaged in four practice mini-games. In each mini-game, subjects were instructed to perform a sequence of six gestures to bring an avatar that was shown on the computer screen from a starting position to a desired goal state (e.g. see Figure~\ref{fig:mini-game}). 
The trial timing epochs (prompting, gesture production, and rest) were as shown in Figure~\ref{fig:YGR-window}. In this block, the classifier model's predicted probabilities were displayed as post-hoc feedback to the user, but were not used to modify the avatar position or state; the avatar always moved one step closer to the goal after each trial, so that each game lasted exactly six moves.
These games were structured so that the $24$ total gestures ($4$ games with $6$ moves each) were evenly distributed among the $8$ active gestures.
After this block, the classification model was retrained from scratch using the labeled data from blocks one and two.
This training set comprised $8$ examples for each of the $9$ classes ($8$ active gestures and ``Rest").

\subsubsection{Block Three: Live Feedback} \label{sec:Block_Three_Live_Feedback}
Only subjects in the veridical feedback and modified feedback groups participated in this block. Subjects performed only one extended trial for each gesture while viewing real-time feedback; in these trials, the gesture production epoch lasted $30$ seconds. Subjects were asked to freely explore their hand posture in order to maximize the predicted probability of the current gesture class, shown on a real-time histogram of the trained model's output. 
For the veridical feedback group, predicted class probabilities were displayed without modification. For the modified feedback group, probabilities were softened towards a uniform distribution as described in Section~\ref{sec:Modified_Feedback}. As discussed previously, the motivation behind this softening procedure was to encourage participants to compensate by performing more precise gestures.
Subjects in the modified feedback group were not informed about this softening procedure.

\subsubsection{Block Four: Free Games} \label{sec:Block_Four_Free_Games}
All subjects were instructed to perform a series of $12$ mini-games. The mini-games had the same structure as in block two, with each game requiring a minimum of six moves to bring the avatar from its starting position to a desired goal state. However, unlike the practice mini-games of block two, subjects were tasked with bringing the avatar to its goal state by planning and performing a gesture sequence of their choice. Critically, the avatar only changed its state when the classifier assigned one class a predicted probability above a decision threshold of $0.5$.
The experimenter manually recorded each attempted gesture to serve as labels for subsequent analysis, and the participant's hand movements were also recorded on video to cross-check these labels.

\section{Signal Modeling} \label{sec:Signal_Modeling}
\subsection{Feature Extraction} As described in Section~\ref{sec:Gesture_Timing}, we extracted raw data for classification from the final $500$ ms of the active gesture production period of each gesture trial.
From each of the $8$ sensor channels of raw sEMG, we computed the Root-Mean-Square (RMS) value and the median frequency of the Fourier spectrum, resulting in $16$-dimensional features.
Given a data vector $x$, RMS is defined as:
\begin{align}
    \textrm{RMS}(x) = \sqrt{ \frac{1}{N}\sum_{i=1}^{N} x_{i}^2}~.
\end{align}
The Median Power Frequency is defined as the frequency value $f_{\textsc{med}}$ that divides the Power Spectral Density (PSD) into two regions with equal power~\cite{hermens1992median}:
\begin{align}
\int_{0}^{f_{\textsc{med}}} \!\textrm{PSD}(f) df = 
    \int_{f_{\textsc{med}}}^{\infty}\! \textrm{PSD}(f) df = 
    \frac{1}{2} \int_{0}^{\infty} \!\textrm{PSD}(f) df ~.
\end{align}

\subsection{Classification Model} \label{sec:Classification_Model}
Given extracted features, we used a two-stage classification pipeline to predict among $9$ possible gestures: Up, Thumb, Right, Pinch, Down, Fist, Left, Open, and Rest.
The classification model consisted of an encoder formed from Support Vector Machine (SVM) models that produced a latent representation, and a logistic regression classifier that produced predicted class probabilities.
In the encoder portion of the model, we trained a one-vs-one (OVO) SVM classifier \cite{Kressel99} for each of the $\binom{9}{2} = 36$ pairs of gestures.
Each of these OVO-SVM models produced a scalar output (representing the probability of assigning to the first of its two classes); these $36$ scalars were stacked into a latent vector and passed to the logistic regression model.

Given a supervised training dataset, we first fit the one-vs-one SVM models using linear programming with the CVXPY Python library \cite{cvxpy}. The linear programming objective we used was based on the semi-supervised SVM formulation of \cite{bennett1998semi}, to allow future semi-supervised extensions. Specifically, the SVM parameters were trained according to the following optimization problem:
\begin{align} \label{eq:svm-obj}
& \min_{w,b,\eta}  
 C \sum_{i=1}^l \eta_i  + 
 \frac{1}{2} \| w \|^2  \\
 \text{s.t.} & \;\;  y_i (w x_i - b) + \eta_i \geq 1, \;\; \eta_i \geq 0, \;\; i = 1,\ldots,l \nonumber  
\end{align}
\noindent where $w,b$ were the parameters to be optimized, $\eta_i$ were slack variables allowing misclassification of individual points, and $C > 0$ is a fixed penalty parameter controlling the margin's strictness.

We implemented the logistic regression classifier with the PyTorch Python library \cite{pytorch} using a single linear layer and a SoftMax function. After the SVM encoder portion of the model was trained, it was held fixed while the logistic regression classifier model was trained by stochastic gradient descent to minimize the cross-entropy loss. We trained the classifier model for $1000$ epochs with a batch size of $20$ and AdamW~\cite{adamw} optimizer. See Algorithm~\ref{alg:training} for a summary of our classifier training procedure.

\paragraph*{Smoothing} As noted, participants in the veridical feedback and modified feedback groups were shown real-time output from the model.
Due to the high sampling frequency of the sEMG sensors used, and the relatively computationally simple prediction model, the system was capable of making very fast adjustments to the predicted output, which can result in unwanted jitter due to slight fluctuations in raw signal or hand positioning.
Therefore, we used an exponential moving average (EMA) to smooth the model's predictions in time.
At time-step $t$, the model produces a raw probability vector $P^{(t)}$, which is then mixed with the previous probability vector using a momentum parameter $\lambda$ to produce a smoothed vector $P_{\textsc{Smooth}}^{(t)}$:
\begin{align}
    P_{\textsc{Smooth}}^{(t)} = \lambda P_{\textsc{Smooth}}^{(t-1)} + (1 - \lambda) P^{(t)}.
\end{align}
For values of $\lambda$ close to $1$, this causes the probability vector to update more slowly and smoothly. We used a value of $0.9$, which alleviated the issue of jitter in the model output, while still allowing model outputs to change quickly between different gestures.

\begin{algorithm}[htb]
\caption{Classifier Training Procedure}
\label{alg:training}
\SetKwInOut{Input}{Input}
\Input{Features $X$, Labels $Y$}
\SetKwInOut{Output}{Output}
\Output{OVO SVM parameters $w$, $b$, Classifier parameters $\theta$}
Initialize $w, b, \theta$ randomly\;
Train $w, b$ on $(X, Y)$ \tcp*[r]{See Eq.~\ref{eq:svm-obj}}
$S \gets \textrm{OVO-SVM}(X)$ \tcp*[r]{SVM scores}
Train Classifier on $(S, Y)$ \;
\Return $w, b, \theta$ \;
\end{algorithm}

\subsection{Modified Feedback} \label{sec:Modified_Feedback} 
As mentioned above, subjects in the modified feedback group were shown modified real-time output from the trained classifier during block three of the experiment.
Specifically, the vector of smoothed predicted probabilities from the model was modified according to the following formula:
\begin{align}
    P_{\textsc{modified}} = \frac{ [P_{\textsc{Smooth}}]^{m} }{\sum\limits_{c \in C} [P_{\textsc{Smooth}}]^{m}}, 
    \label{eqn:modify-probs}
\end{align}
where the modification exponent $m$ was set to $0.75$, and $C$ represents the $9$ classes used.
The value of $m$ was chosen subjectively to make a noticeable effect while not being too extreme; since subjects must still be able to exceed a decision threshold of $0.5$ for a gesture to be correct.

Note that this feedback can be viewed as a form of error augmentation. When asked to perform a certain target gesture, we can consider the error to be the distance (e.g. cross-entropy distance or L2 norm) between the model's predicted probability vector and an idealized probability vector in which all mass is concentrated on the target class. Subjects in both feedback groups were instructed to explore gestures and maximize the predicted probability of the target class; thus they were instructed to minimize this error. However, subjects in the modified feedback group viewed a flattened probability vector; this flattening causes the vector to appear to have greater error. See Figure~\ref{fig:live-feedback} for an example.

\subsection{User Interface and Software Design} \label{sec:User_Interface_and_Software_Design}
\begin{figure}[htb]
    \centering
    \includegraphics[width=1.0\linewidth,clip,trim=0 0 0 30]{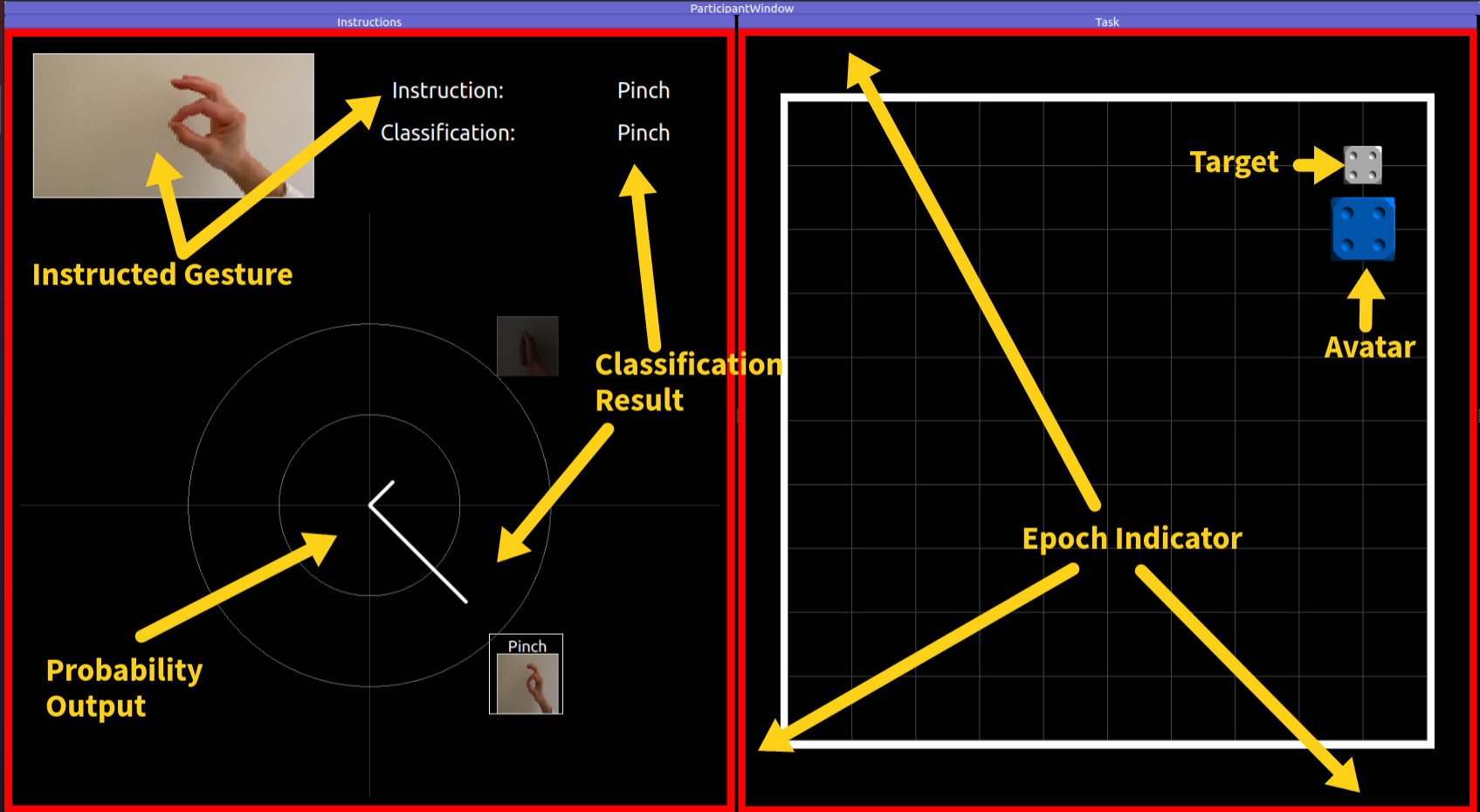}
    \caption{The participant User Interface. Top left: instructed gesture. Bottom left: predicted gesture probabilities. Right: Task window including subject's avatar and target. Outer edge: gesture epoch indicator.}
    \label{fig:participant-window}
\end{figure}

Figure~\ref{fig:participant-window} shows the user interface (UI) displayed to participants. All components of the UI were implemented using PyQt Python package \cite{pyqt_docu}.
Data collection and real-time processing were performed using the LabGraph Python package \cite{labgraph2021}. 
On the top left, the UI displayed an instructed gesture via image and text during blocks one and two (see Section~\ref{sec:Block_One_Calibration} and \ref{sec:Block_Two_Instructed_Games}).
On the bottom left, the UI showed post-hoc predicted probabilities for each gesture as a radial plot. 
The length of each line was scaled according to the value; the outer circle represented a value of $1$, and the inner circle represented a value of $0.5$ (i.e. the model's decision threshold).
The opacity of gesture images around the radial plot was also scaled according to the value.
The outer edge of the UI was colored yellow, green, or red to indicate gesture timing epoch as described in Section~\ref{sec:Gesture_Timing}.
On the right of the UI was the task window in which the mini-games were played during blocks two and four (see Section~\ref{sec:Block_Two_Instructed_Games} and \ref{sec:Block_Four_Free_Games}).
As described previously, participants used one of $8$ active gestures to move their avatar (the blue die). 
The goal of each mini-game in blocks two and four was to use these gestures to match the blue die to the gray target die.

\begin{figure}[htb]
    \centering
    \includegraphics[width=0.8\linewidth,clip,trim=0 0 0 60]{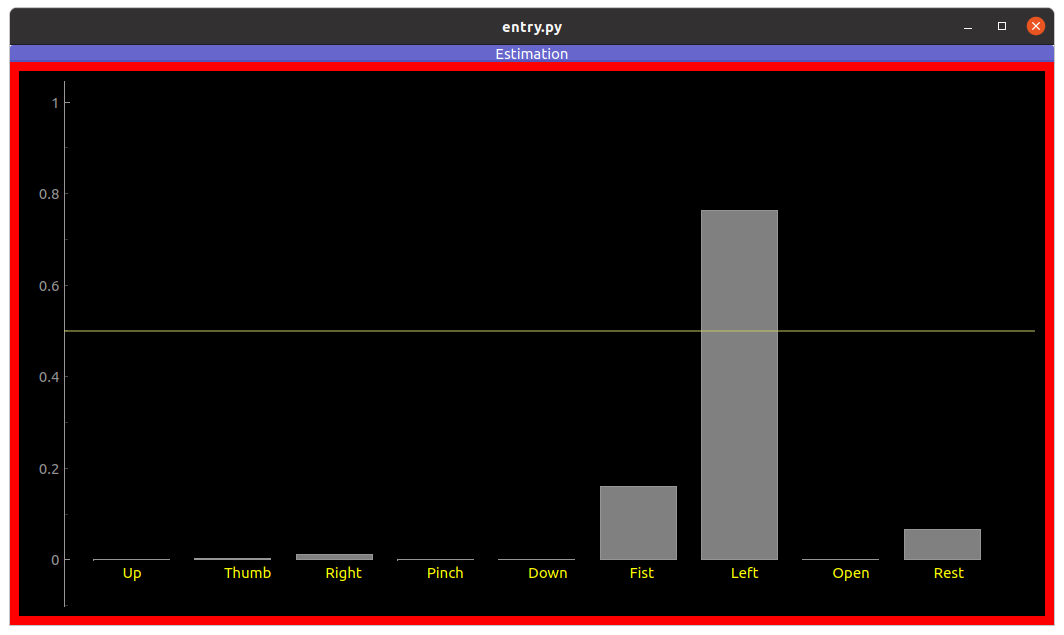}
    \includegraphics[width=0.8\linewidth]{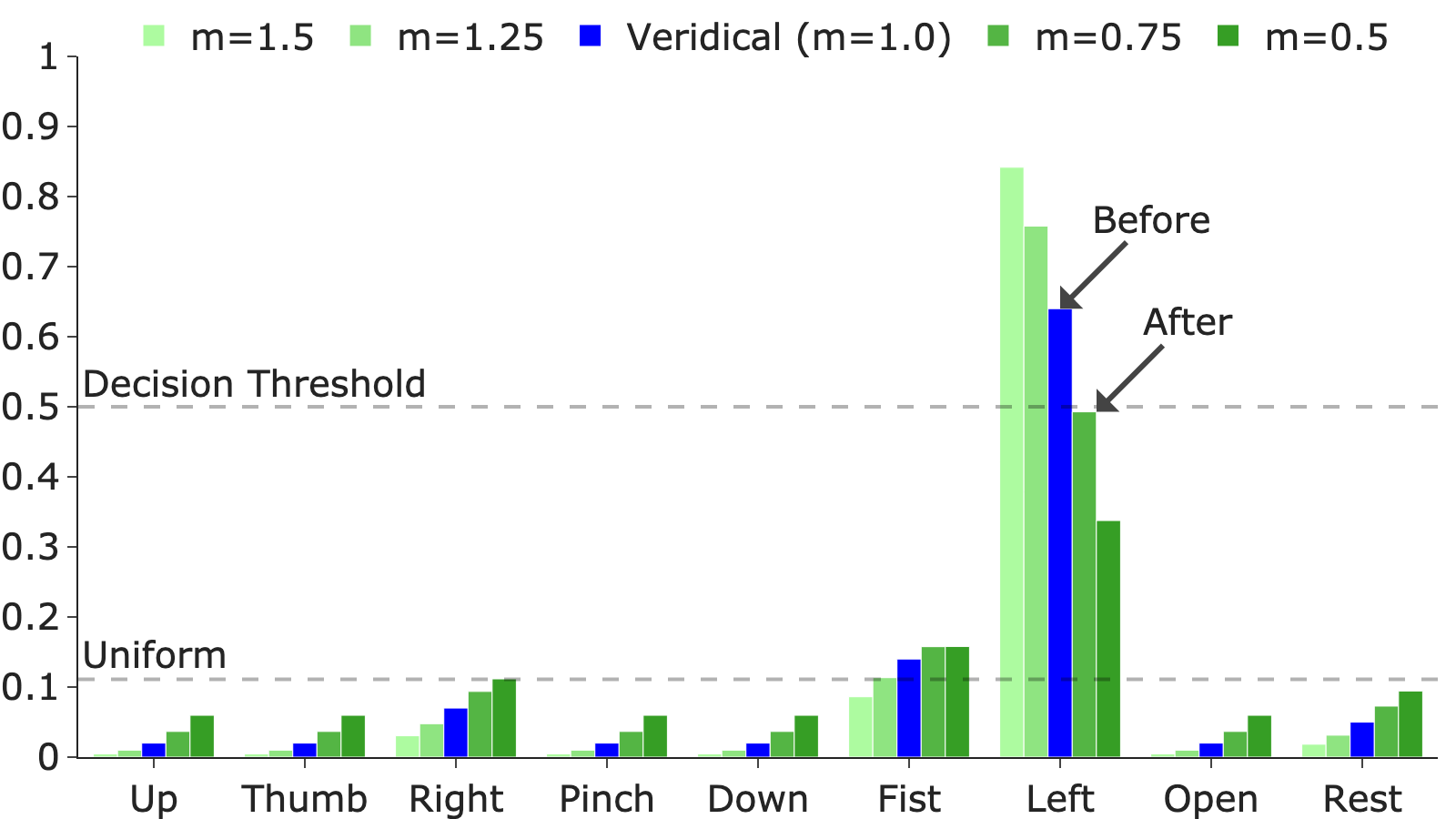}
    \caption{Top: Real-time probability feedback window. The horizontal line at $0.5$ shows the decision threshold. Bottom: Example of probability values without modification (``Veridical'') and with modification (``Modified'') as described in Sec.~\ref{sec:Modified_Feedback} for several hypothetical values of $m$. $m=0.75$ used for real experiments. Arrows highlight an example case where modification causes the gesture to become sub-threshold; participant may compensate by improving gesture quality.}
    \label{fig:live-feedback}
\end{figure}

\paragraph{Error Augmentation in Live Feedback} During block three (see Section~\ref{sec:Block_Three_Live_Feedback}), participants who received real-time feedback were presented with a different display, as shown in Figure~\ref{fig:live-feedback}.
Here, the probability of each class was displayed using a bar plot that was updated in real-time.
The participant's goal during this block of the experiment was to explore hand positions in order to maximize the predicted probability of the current gesture class. For participants in the modified feedback group, model outputs were flattened towards a uniform distribution using Equation~\ref{eqn:modify-probs}.

\subsection{Classifier Metrics}
As mentioned in Section~\ref{sec:Block_Four_Free_Games}, the experimenter recorded each intended gesture made by the participant, so that model accuracy could be evaluated after-the-fact.
Accuracy was defined as the fraction of correctly classified items.
In addition to the $8$ active gestures and the ``rest" class, the decision threshold of $0.5$ that was used resulted in another possible outcome for gesture trials when no gesture rose above the decision threshold, which we refer to as ``NoClass".
Gesture trials in which the subject was not prepared to make a gesture during the ``gesture production" epoch were recorded as having a true label of ``Rest".

\subsection{Feature-Space Class Structure} \label{sec:Feature_Space_Class_Structure}
To evaluate how feedback affects human learning, we analyzed the feature-space distribution of trials from different gestures performed in block four of the experiment. This feature-space representation does not depend on the model, since these features are obtained using simple, deterministic transformations of the raw data (RMS and median frequency after Fourier transform). The differences in feature-space class structure across treatment groups can therefore give information about human learning.

Previous research has introduced a variety of feature space metrics for similar tasks, such as separability index and repeatability index \cite{franzke2020exploring, bunderson2012quantification}.
Such metrics are based on the Mahalanobis distance and require computing a class covariance matrix. Since our experiment is focused on short calibration times and we operated in a regime of limited data, we do not have enough samples to compute reasonable estimates of class covariance matrices, even with shrinkage techniques. We therefore used feature-space metrics based on pairwise comparisons between samples. 

\paragraph{Kernel Similarities}
We base our analysis of feature-space structure on a Radial Basis Function (RBF) kernel similarity measure. The RBF kernel computes a similarity measure that corresponds to an implicit infinite-dimensional vector space. For two feature vectors $x, x'$ belonging to a dataset $X$ and a length scale parameter $\gamma \in \mathbb{R}$, the RBF kernel similarity is computed as:
\begin{align}
    RBF(x, x', \gamma) & = \exp \left(- \gamma \| x - x' \|^2 \right).
\end{align}
The length scale $\gamma$ is an important hyperparameter that determines the rate at which similarities decay as two points are moved farther apart. 
We follow the so-called ``median heuristic" \cite{median-heuristic}, in which $\gamma$ is set based on the median length scale of a dataset $X$:
\begin{align}
    \gamma_{\textsc{med}} & = 1 / \textrm{med}(\|x - x'\|^2 , \ \forall \ (x, x') \in \{X \times X \}).
\end{align}
We set $\gamma_{\textsc{med}}$ individually for each subject, based on all of their pooled gesture trials.

Note that this approach is effectively a non-linear rescaling of pairwise Euclidean distances, and also handles the potential issue of outlier points having extremely large Euclidean distances.

\paragraph{Class Similarity Matrices} 
We use this notion of kernel similarity to construct a class similarity matrix for each subject. For classes $C_1, \ldots, C_{\mathscr{C}}$, we build a square, symmetric matrix $D \in \mathbb{R}^{(\mathscr{C} \times \mathscr{C})}$ such that the entry at position $(i, j)$ describes the average RBF kernel similarity between items in classes $C_i$ and $C_j$:
\begin{align}
    D_{ij} & = \frac{1}{|C_i| |C_j|} \sum_{x \in C_i} \sum_{x' \in C_j}  RBF(x, x', \gamma_{\textsc{med}}). \label{eqn:similarity_matrix}
\end{align}
After computing the entries in a similarity matrix, we normalize the entries to the range $[0, 1]$ so that these matrices may be easily compared across subjects and groups.

\begin{figure}
    \centering
    \includegraphics[width=0.59\linewidth]{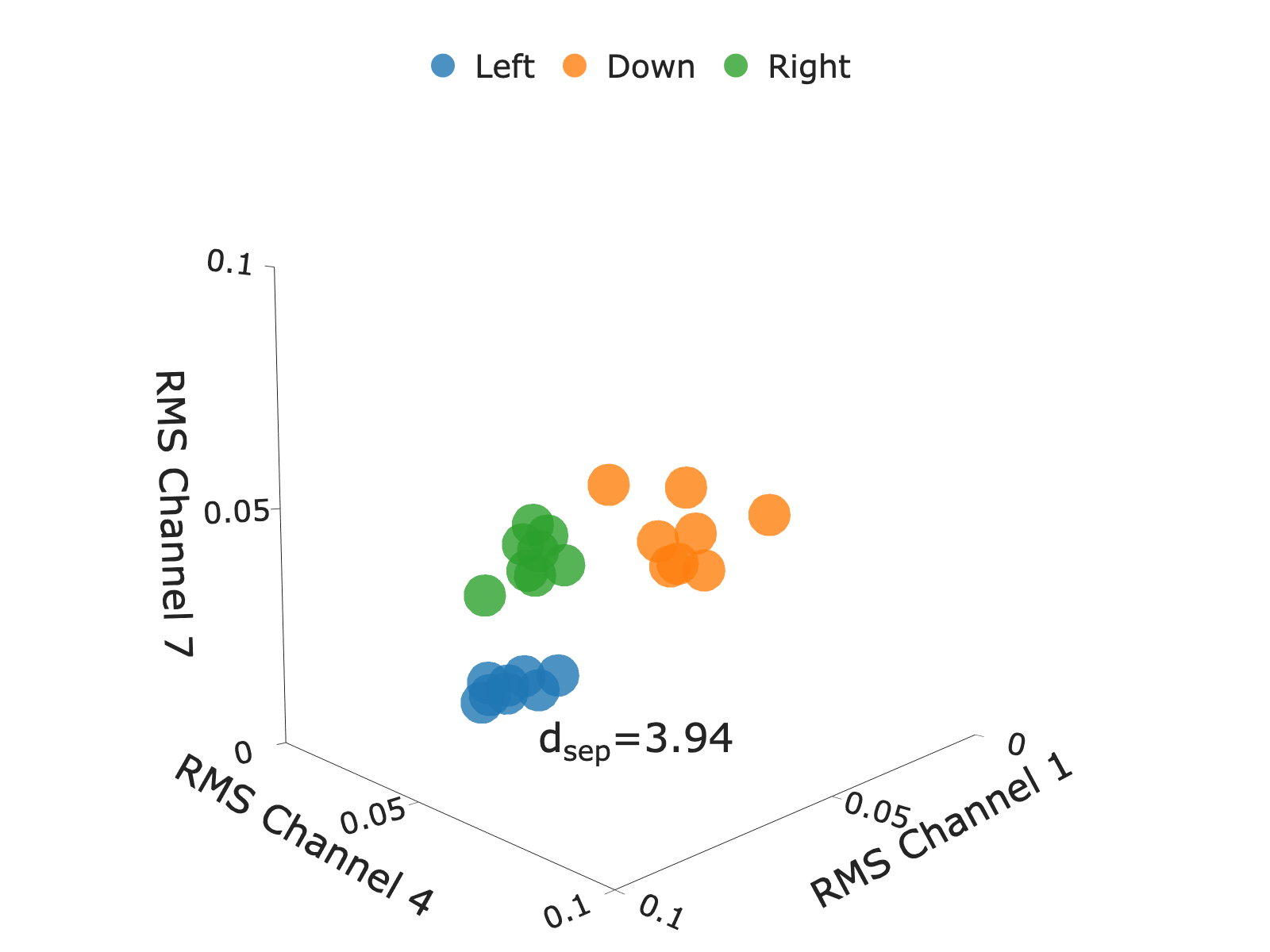}    
    \includegraphics[width=0.39\linewidth]{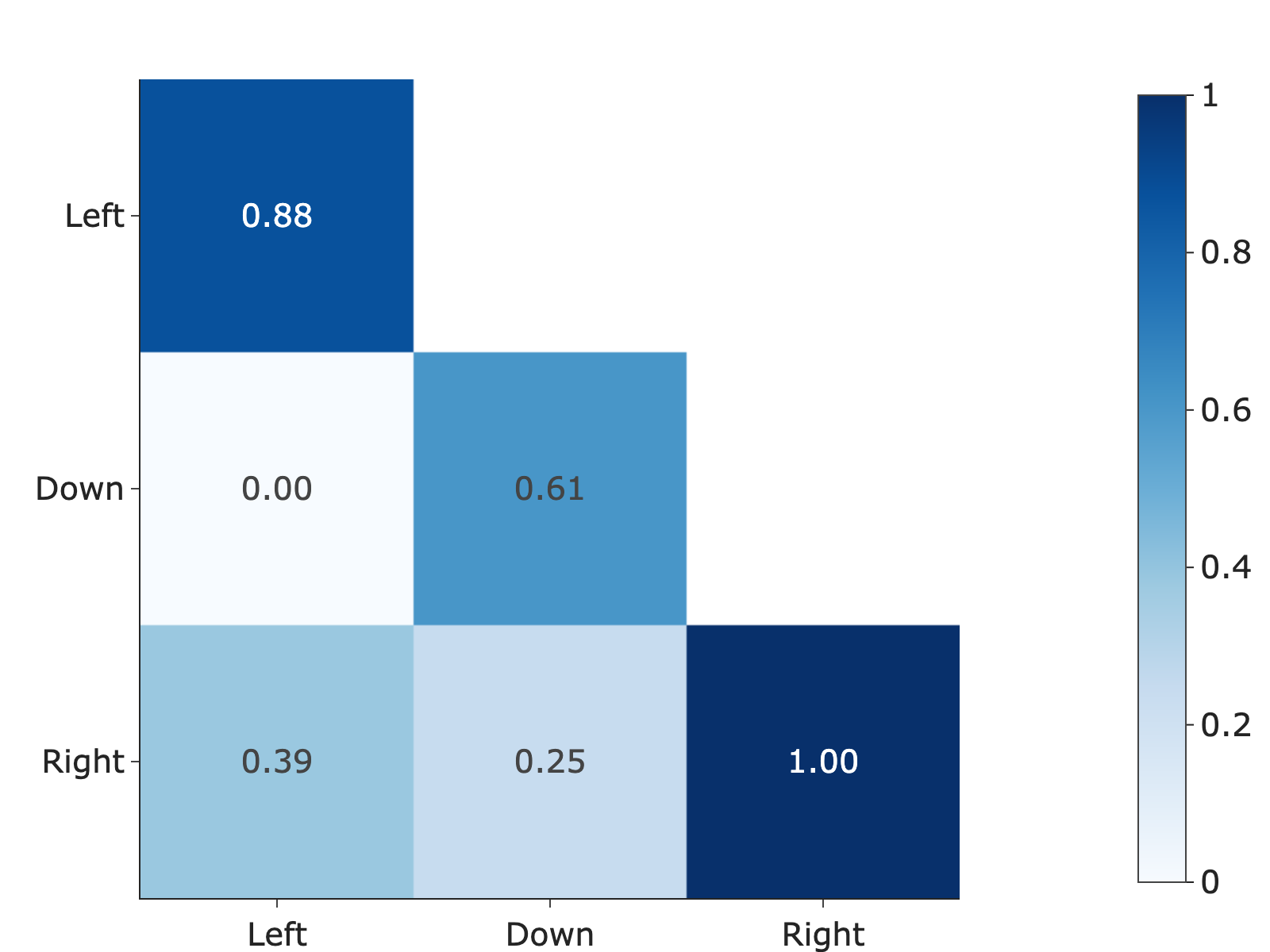}
    \includegraphics[width=0.59\linewidth]{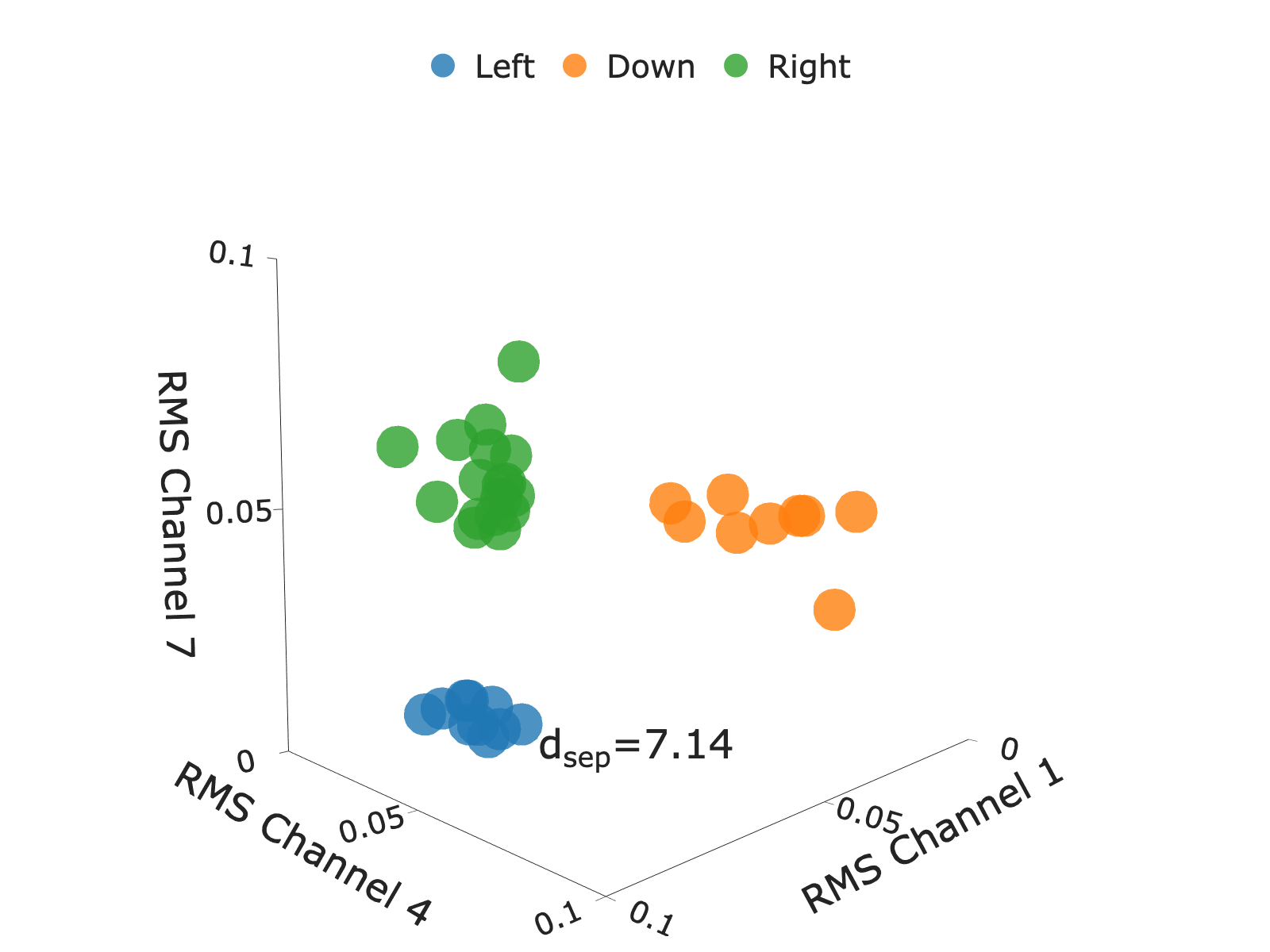}
    \includegraphics[width=0.39\linewidth]{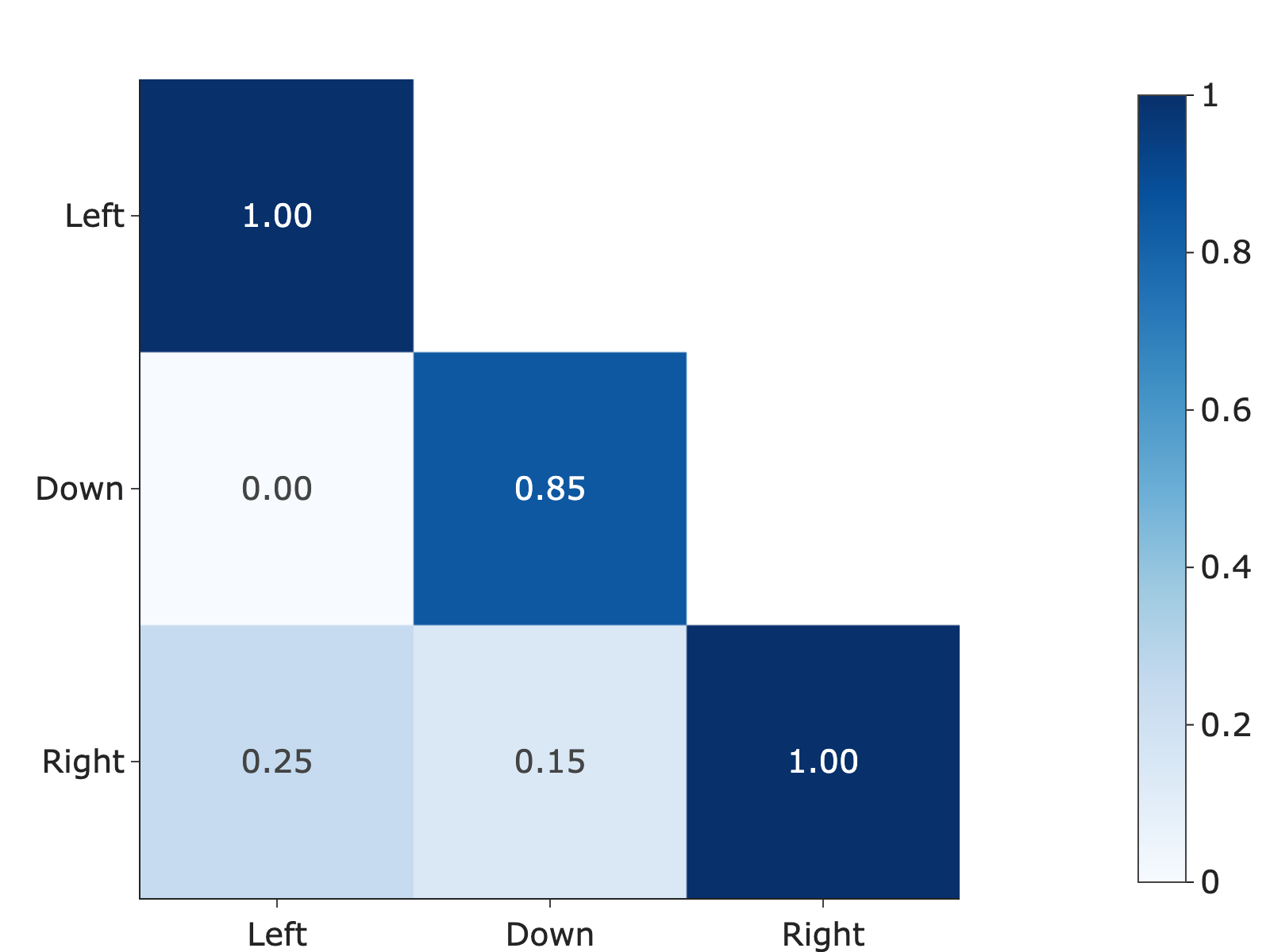}
    \caption{Didactic example for class similarity matrices $D$ and scalar class separation measure $d_{\textsc{sep}}$. For a chosen subject from the Modified condition, we analyze $3$ of the original $16$ features (RMS value from electrodes 1, 4, and 7) and a subset of gestures (``Left'', ``Down'', and ``Right''). Top row: features from calibration and instructed blocks. Bottom row: features from free games. Left: Scatter plot of $3$-dimensional features, and scalar class separation value. Right: The corresponding class separation matrix.}
    \label{fig:example-similarities}
\end{figure}

Classes that are closer together in feature space will have a higher average similarity and therefore a larger entry in this similarity matrix.
A subject whose gestures are easily classifiable may tend to have precise gestures that are also well-separated from each other.
This would result in having a high average similarity between trials in the same gesture class (diagonal entries of the class similarity matrix) and a low average similarity between trials of different classes (off-diagonal entries).
See Section~\ref{sec:Class_Similarity_Matrices} for class similarity matrices from each experimental group, and see Figure~\ref{fig:example-similarities} for didactic examples of similarity matrix $D$.

\paragraph{Scalar Class Separation Measure}
In order to look for trends in the feature-space distribution over time and to identify global trends across groups, we also summarize these normalized class similarity matrices using a scalar class separation measure, $d_{\textsc{sep}}$, which we define as the average within-class similarity divided by the average between-class similarity. Given a normalized similarity matrix $D$ as described above,
\begin{align}
    d_{\textsc{sep}} & = 
        \left( \frac{1}{N} \sum_{i=1}^N D_{ii} \right) /
        \left( \frac{2}{N (N-1)} \sum_{i = 2}^{N} 
        \sum_{j=1}^{i-1} D_{ij} \right). \label{eqn:scalar_separation}
\end{align}
As indicated above, larger within-class similarities indicate that trials from the same gesture are precise and repeated with high-fidelity, while smaller between-class similarities indicate that trials from different gestures are easily distinguished.
Thus, a dataset with a larger value of $d_{\textsc{sep}}$ may contain gestures that will be more easily classified.

In Figure~\ref{fig:example-similarities}, we show examples of class similarity matrix $D$ and scalar similarity measure $d_{\textsc{sep}}$. To produce an example that can be easily visualized, we select a subject from the ``Modified'' condition that showed a large improvement in feature-space separation.
For this subject, we select three gestures (``Left'', ``Down'', and ``Right'') and three features (RMS value from electrodes 1, 4, and 7). In the top row, we show metrics for this subject's data during the ``Calibration'' and ``Instructed'' blocks, and in the bottom row, we show metrics from the ``Free'' block; recall that the subject experiences live feedback training after the ``Instructed'' block. 
We observe that the features of each class become more distinct after the user performs live feedback training; this is captured as an increase in the similarities on the diagonal of $D$ and a decrease in similarities off-diagonal. These changes in $D$ are also summarized in $d_{\textsc{sep}}$, which increases from $2.8$ to $3.55$.

\subsection{Within-Subject Normalization} \label{sec:Within_Subject_Baseline_Subtraction}
The focus of this work is to measure the effect of the proposed veridical and modified feedback strategies on subject performance.
We note that overall subject performance may be influenced by a relatively large number of factors of variation, such as factors affecting dexterity and motor precision, subject motor learning speed, and subject-intrinsic factors affecting raw sEMG signal-to-noise ratio.
Thus, a prohibitively large sample size may be required to account for this variation without normalization.
We therefore adopt a within-subject normalization strategy, obtaining baseline statistics for each subject using only data measured \textit{before} our interventions.

For each subject, we measure baseline accuracy by training a model from scratch using that subject's block one data (calibration, Section~\ref{sec:Block_One_Calibration}), and testing this model's classification accuracy on the subject's block two data (instructed games, Section~\ref{sec:Block_Two_Instructed_Games}). 

We obtain baselines for class similarity matrices in the same manner. Within each subject, we collect all gesture trials from the first two experimental blocks, and compute a normalized class similarity matrix. This is subtracted from the matrix computed using data from block four (free games, Section~\ref{sec:Block_Four_Free_Games}) to visualize the difference in similarity for each class. Note that due to the short experimental design, we have relatively few samples per class with which to construct each matrix, and therefore this representation may be somewhat noisy. 

We transform the normalized similarity matrix describing blocks one and two into the scalar class separation measure $d_{\textsc{sep}}$, and likewise transform the similarity matrix describing block four. This results in a baseline-subtracted class separation measure.

Overall, we measure changes from baseline as follows:
\begin{align}
    \Delta \textrm{Acc} & = \textrm{Acc}_{\textsc{Free}} - \textrm{Acc}_{\textsc{Baseline}} \label{eqn:changes_from_baseline} \\
    \Delta D & = D_{\textsc{Free}} - D_{\textsc{Baseline}} \nonumber \\
    \Delta d_{\textsc{sep}} & = d_{\textsc{sep, free}} - d_{\textsc{sep, baseline}} \nonumber
\end{align}

\subsection{Statistical Analysis}
We performed several statistical analyses to determine the effect of feedback on classification accuracy and feature space class separation. Differences between feedback groups at baseline ($\textrm{Acc}_{\textsc{Baseline}}$, $d_{\textsc{sep, baseline}}$) were analyzed using one-way ANOVAs. Likewise, the effect of the feedback group on change scores ($\Delta \textrm{Acc}$, $\Delta \textrm{D}$) was analyzed with one-way ANOVAs ($\alpha = 0.05$). Alpha level was set at 0.05. Significant findings were further analyzed using post-hoc paired comparisons with Bonferroni correction for multiple comparisons. One-sided one-sample t-tests with Bonferroni correction for multiple comparisons ($\alpha = 0.0167$) were used on change scores to test whether each feedback group significantly increased accuracy and distance.  

\section{Results} \label{sec:Results}

All participants were able to successfully complete the experiment, with no reported adverse events. 

\subsection{Group Baselines}
\label{sec:baselines}
\begin{figure}[htb]
    \centering
    \includegraphics[width=\linewidth]{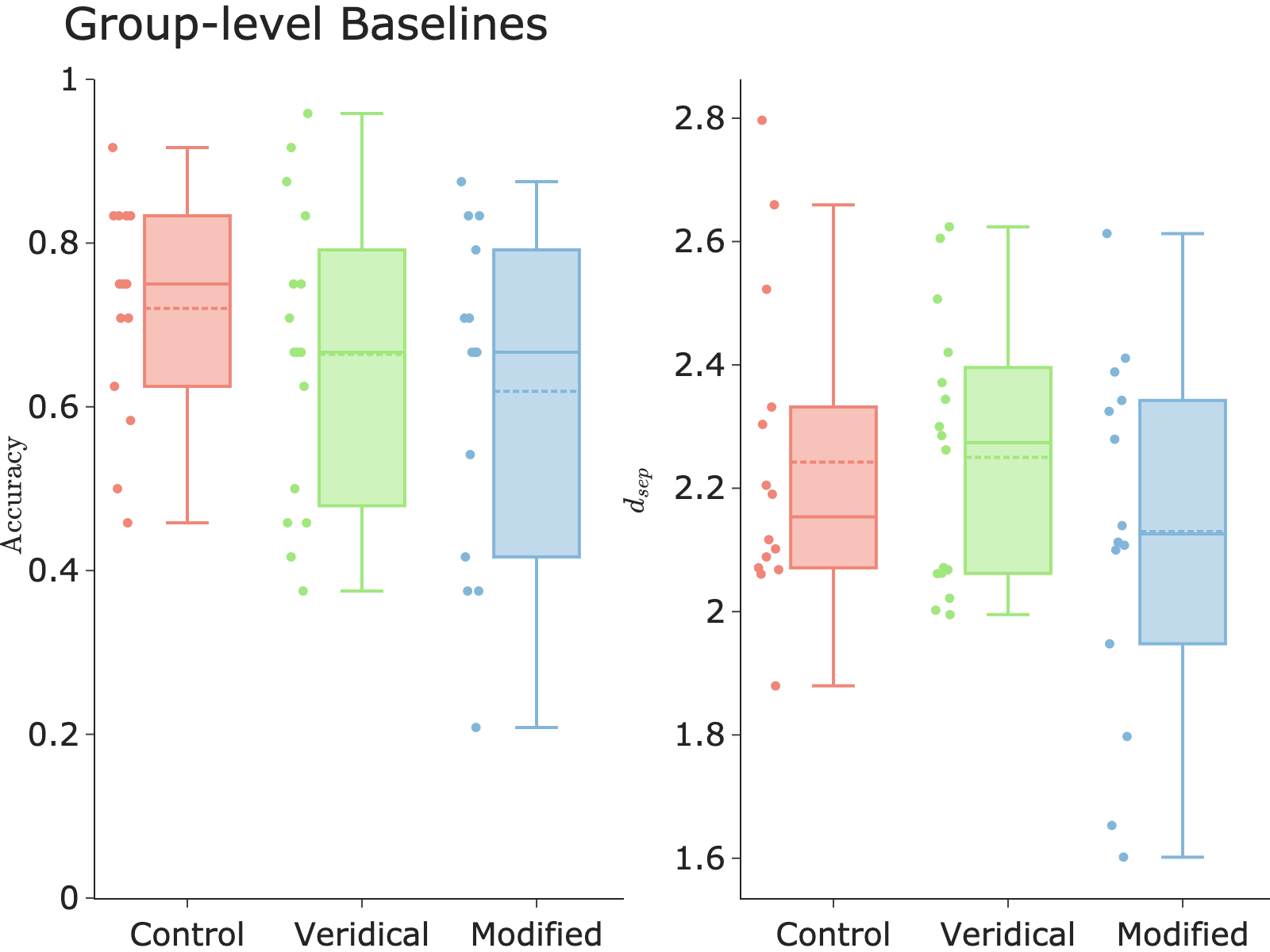}
    \caption{Baseline Performance. Left: Accuracy. Right: Scalar class separation measure $d_{\textsc{sep}}$. Boxplots show the median and quartiles; dotted lines show the mean. Note the relative difference in subject baseline task performance, visible as a gap in baseline accuracy. This discrepancy (due to random group assignment and low subject number) indicates the need for within-subject normalization, as described in Section~\ref{sec:Within_Subject_Baseline_Subtraction}. See Section~\ref{sec:baselines} for statistical analysis.}
    \label{fig:group_baselines}
\end{figure}

In order to check whether random group assignment was a potential confounding factor in our comparisons between groups, we analyzed baseline metrics for each experimental group.
One-way ANOVA indicated no significant differences in baseline accuracy ($F(2,43) = 1.15$, $P = 0.326$) or class separation ($F(2,43) = 0.99$, $P = 0.380$) between experimental groups. 
Figure~\ref{fig:group_baselines} shows a group-level summary of the baseline accuracy and class separation measure. Though no significant differences were found, mean baseline accuracy and class separation scores were greatest in the Control group and smallest in the Modified group.  

\subsection{Effects of Feedback}
\label{sec:feedback}
Individual one-sided one-sample t-tests were used to test for significant improvement in Free block performance from baseline (Bonferroni corrected for 3 comparisons, $\alpha = 0.0167$). For accuracy, only the Modified group showed significant improvement ($t(13) = 2.566$, $P = .012$). No group showed a significant improvement in class separation. One-way ANOVAs indicated no significant between-group differences in $\Delta \textrm{Acc}$ ($F(2,43) = 0.413$, $P = 0.665$) or $\Delta d_{\textsc{sep}}$ ($F(2,43) = 2.022$, $P = 0.145$). 

\begin{figure}[htb]
    \centering
    \includegraphics[width=\linewidth]{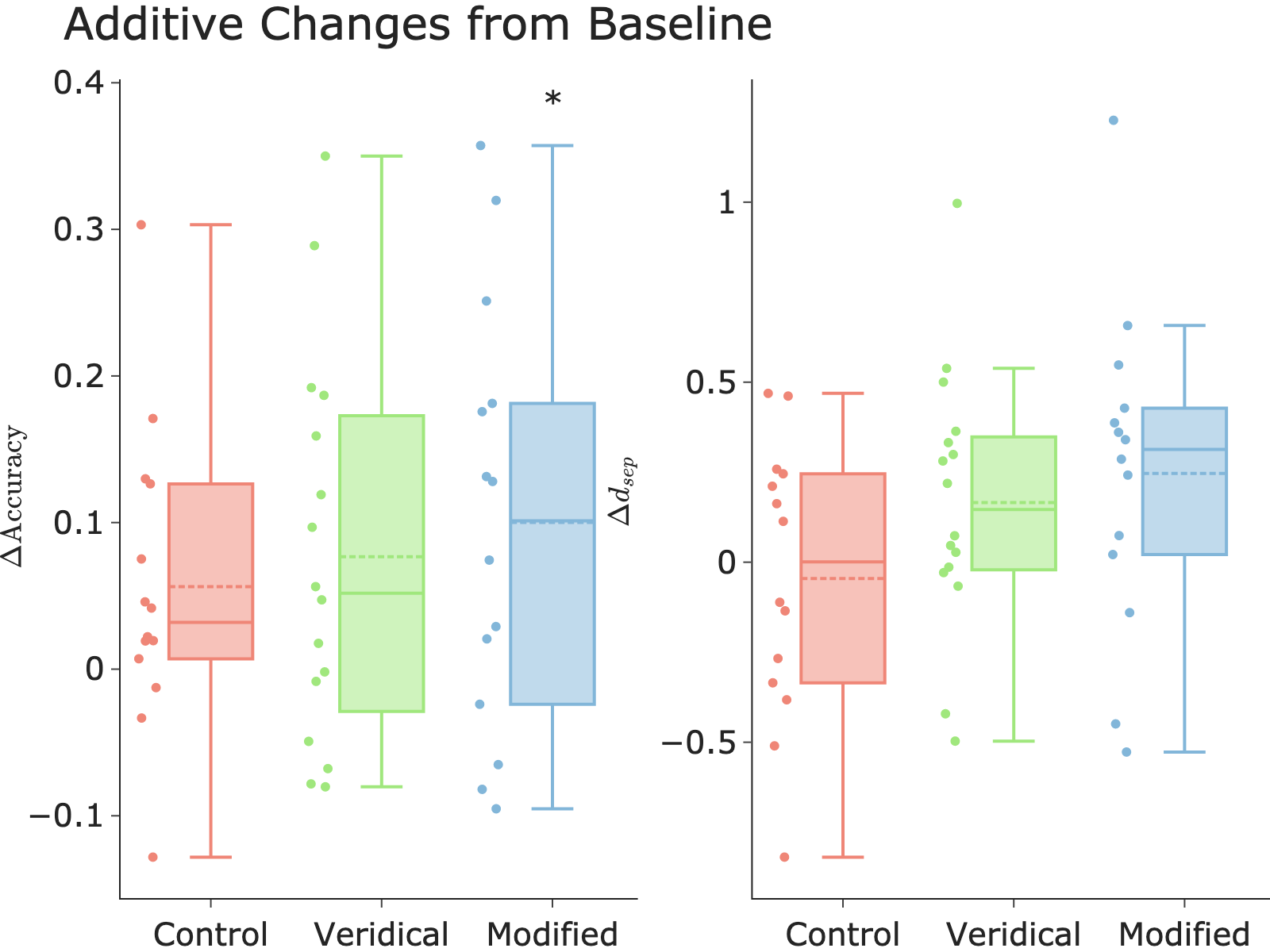}
    \caption{Overall Changes from Baseline Performance. Left: Change in accuracy. Right: Change in scalar class separation measure $d_{\textsc{sep}}$.
    Boxplots show median and quartiles; dotted lines show mean.
    For each subject, we perform baseline subtraction as described in Section~\ref{sec:Within_Subject_Baseline_Subtraction}.
    Change in accuracy for the modified group was significantly greater than zero using; see Section~\ref{sec:feedback} for statistical analysis.
    }
    \label{fig:group_changes}
\end{figure}

Figure~\ref{fig:group_changes} shows the average change from baseline performance in each experimental group, as measured in the accuracy of gesture classification (left panel) and feature-space class separation measure (right panel). 
These data demonstrate that, on average, the increase in performance over the course of the experiment was greatest for subjects in the modified feedback group.
Note that the variation between subjects is relatively high, resulting in overlapping estimates of mean performance.
We observe that both groups that received real-time feedback exhibited larger variation; in particular, the interquartile range for these two groups ($0.18$ and $0.19$ units for Veridical and Modified, respectively) is nearly twice the range of the control group ($0.10$ units). This may indicate that some subjects are better at learning from this form of visual feedback than others, or that some subjects were adversely affected by feedback while others were positively affected.

\subsection{Class Confusion}
\label{sec:confusion_matrices}
\begin{figure*}
    \centering
    \includegraphics[width=0.32\linewidth]{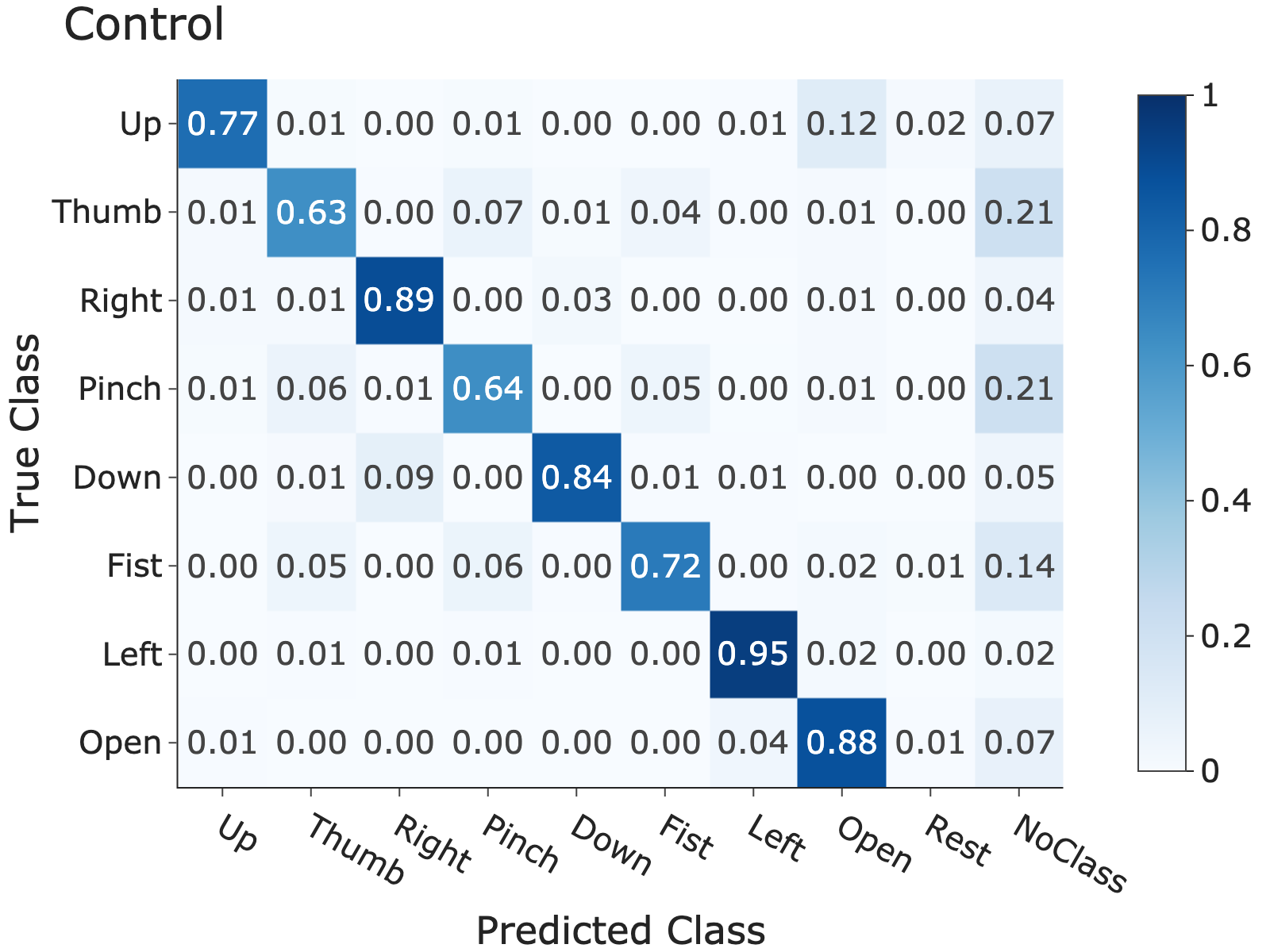}
    \includegraphics[width=0.32\linewidth]{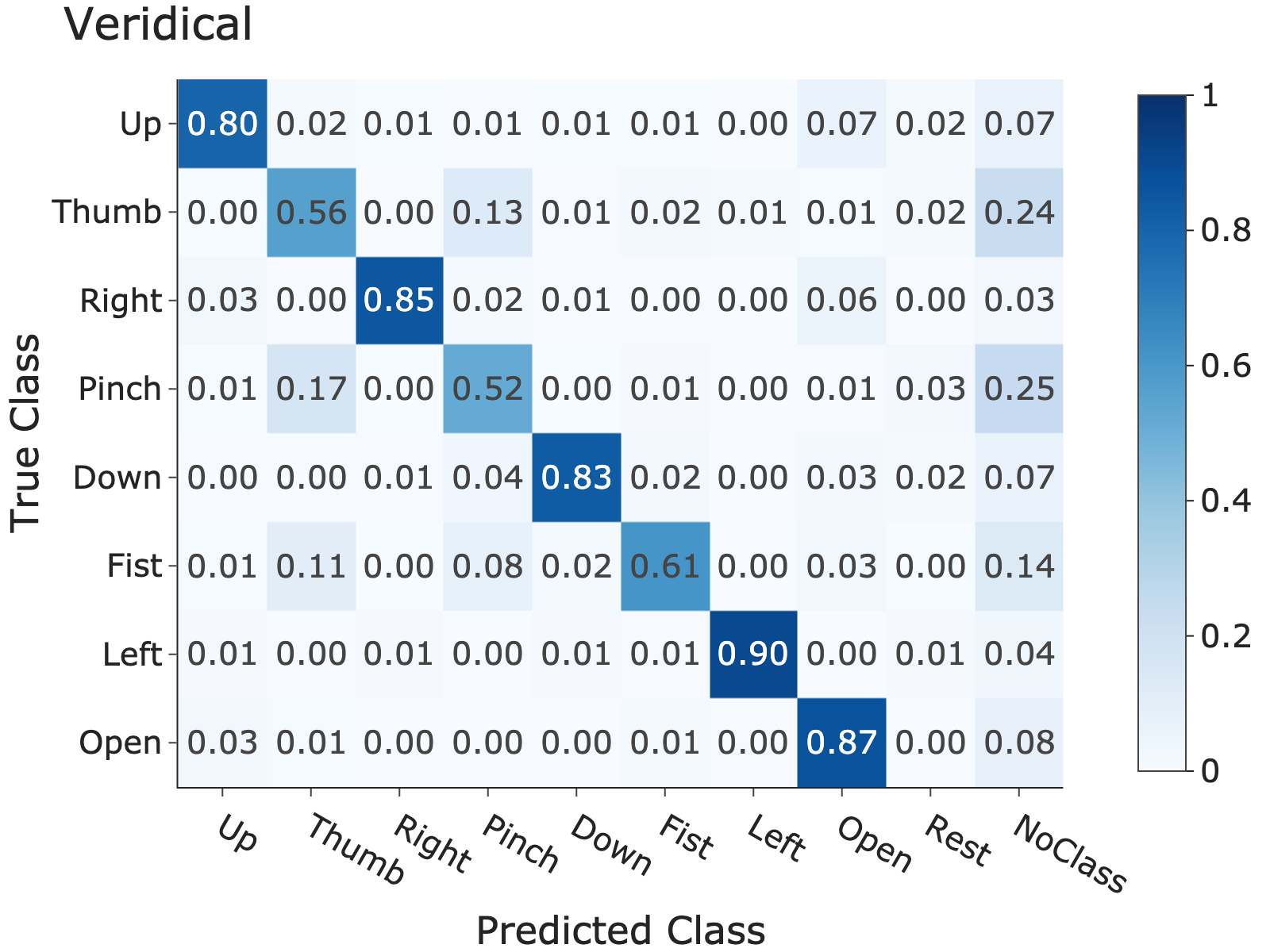}
    \includegraphics[width=0.32\linewidth]{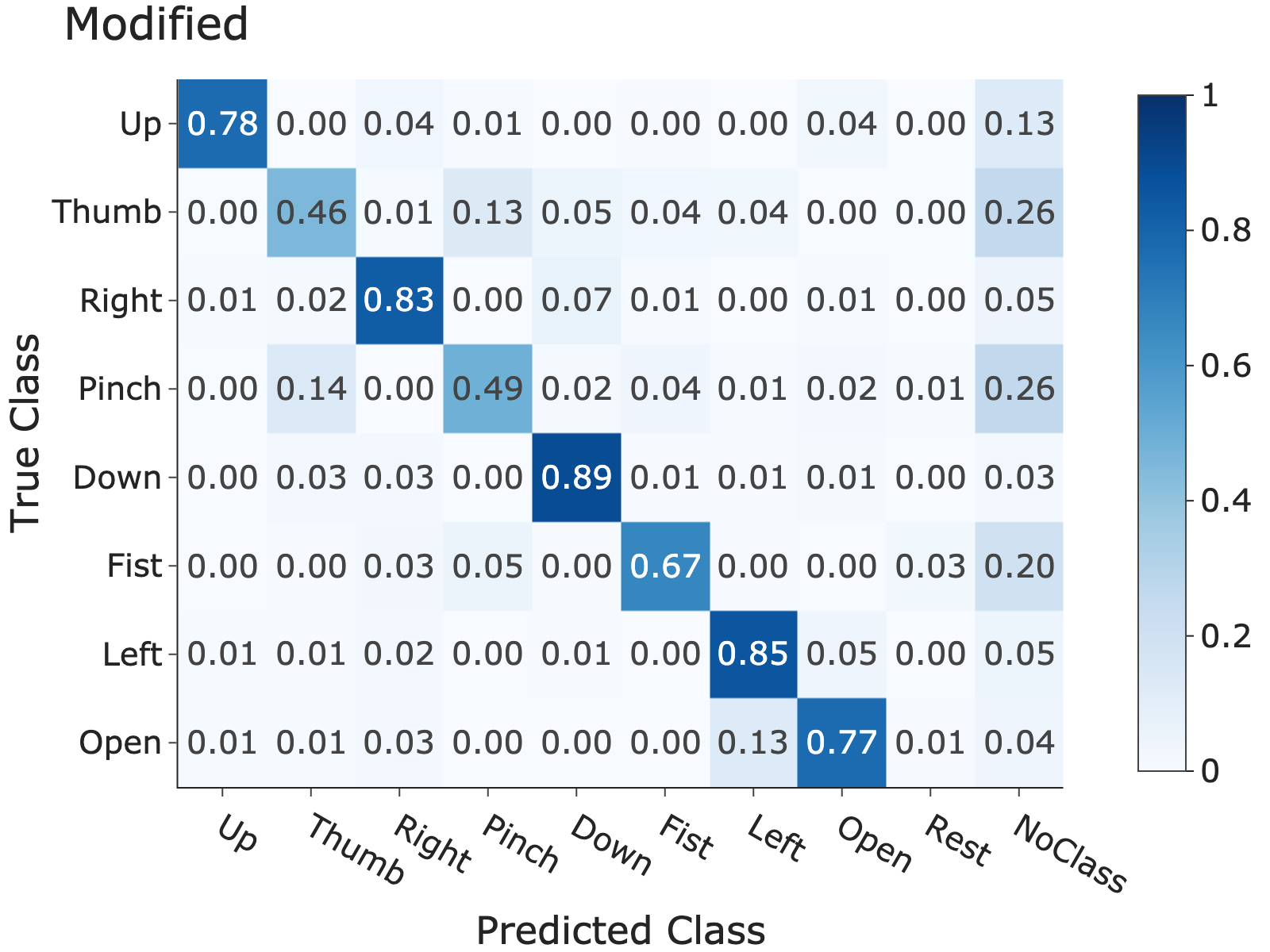}
    \caption{Confusion Matrices averaged across subjects and normalized within each row. No within-subject correction is applied. Class confusion structure is largely similar across groups. Left: Control subject. Middle: Veridical feedback. Right: Modified Feedback.}
    \label{fig:group_conf_mats}
\end{figure*}
Figure~\ref{fig:group_conf_mats} shows the group average confusion matrices of gesture trials during block four (free games) for each group. Rows represent the classification of the attempted gesture, normalized to $1$. 
There are notable similarities across the groups, indicating several gestures that are intrinsically difficult and gesture pairs that are inherently close. In particular, the ``thumb", ``pinch", and ``fist" gestures all have a large fraction (about $25\%$) of gestures that fall below the decision threshold. Similarly, there was an overall trend that these three gestures tended to be confused, resulting in non-zero entries for the off-diagonal entries (fist, thumb), (fist, pinch), (thumb, pinch), etc. 
The similarity between groups is an indication that feedback did not grossly disrupt subject behavior for certain gesture classes or cause substantially different effects for different classes.

\subsection{Class Feature Space Similarity} \label{sec:Class_Similarity_Matrices}
\begin{figure*}[htb]
    \centering
    \hspace{-3.5mm}
    \includegraphics[width=0.30\linewidth,clip,trim={0 0 350 0}]{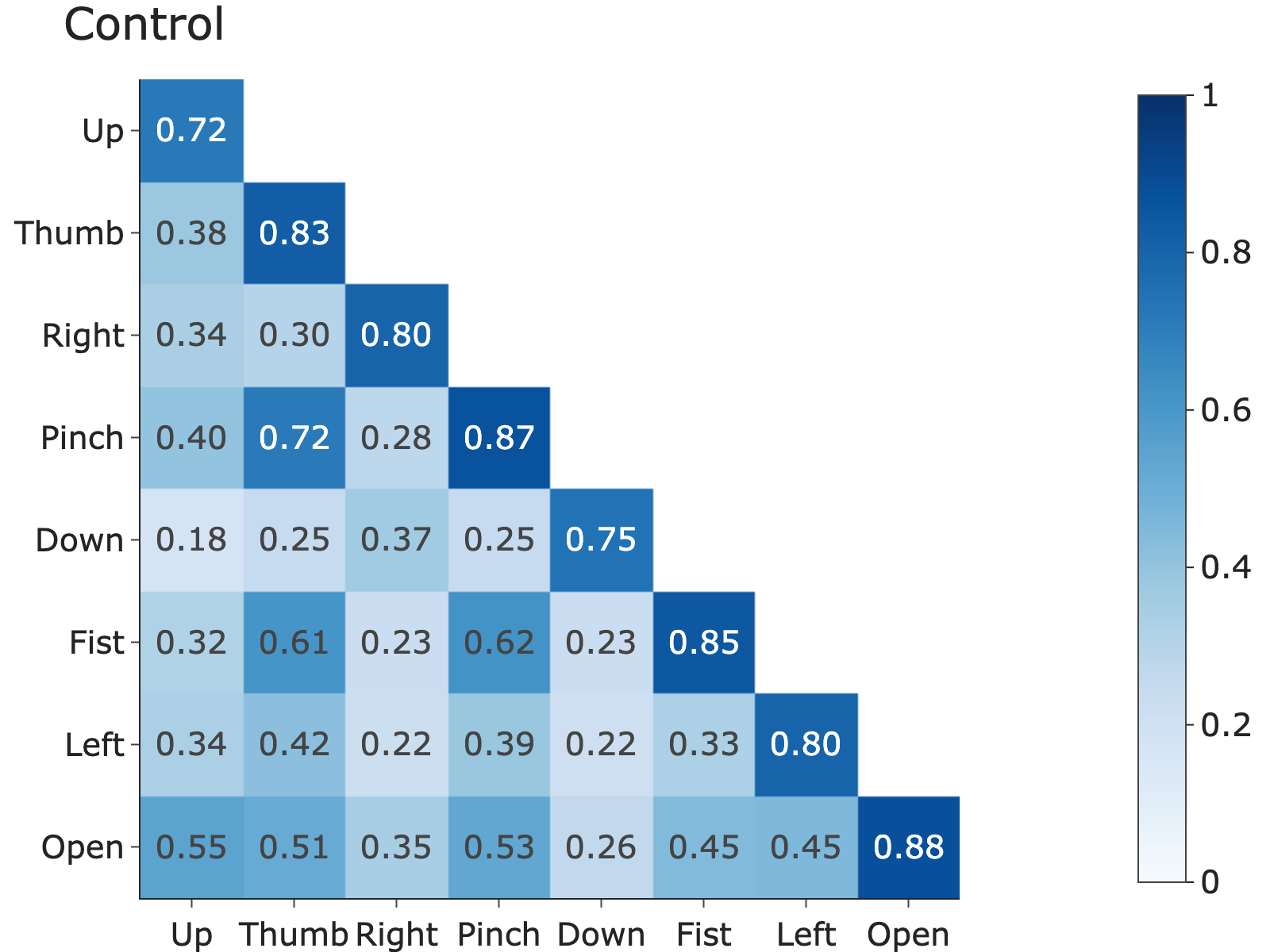}
    \includegraphics[width=0.30\linewidth,clip,trim={0 0 350 0}]{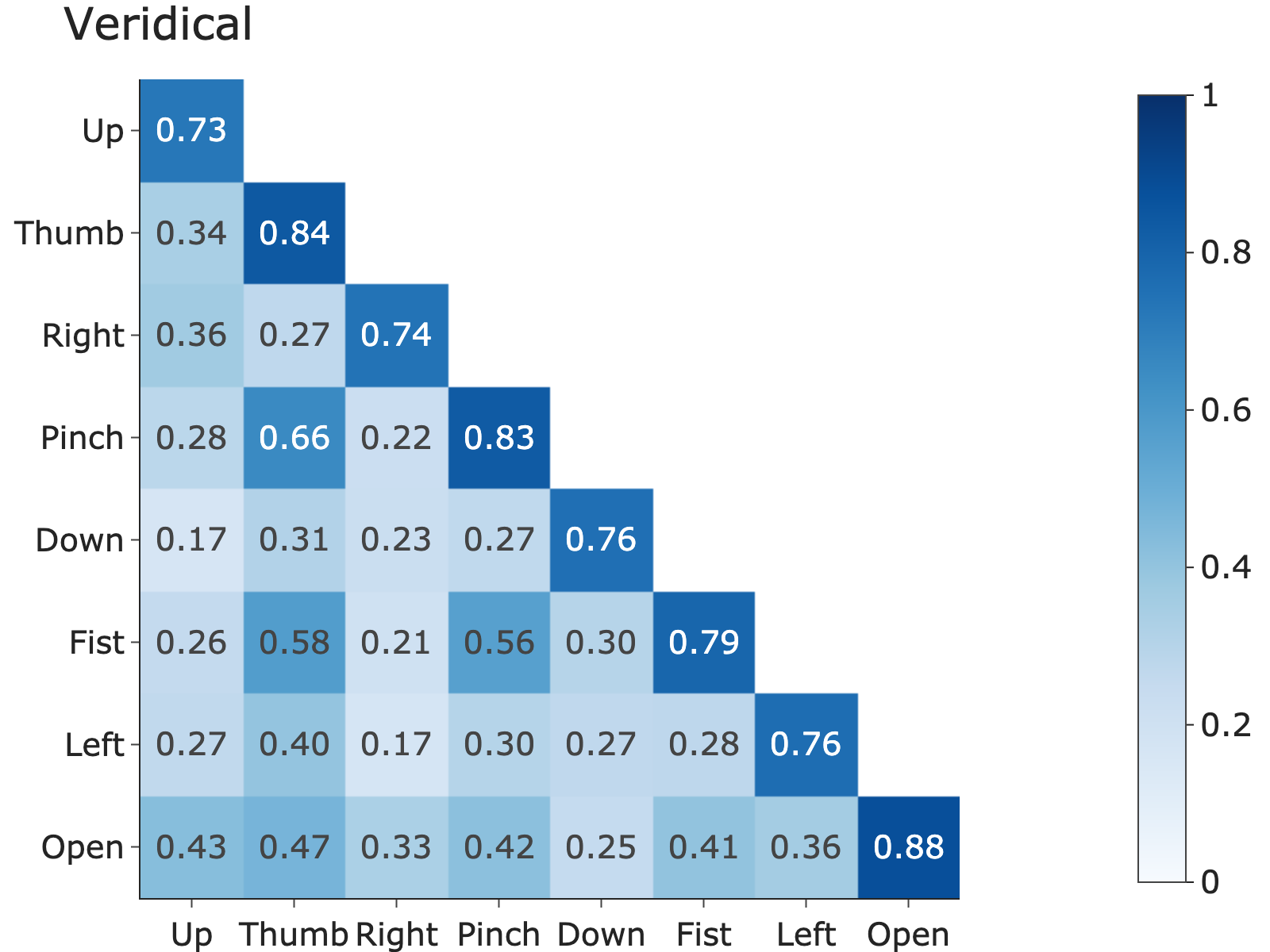}
    \includegraphics[width=0.30\linewidth,clip,trim={0 0 350 0}]{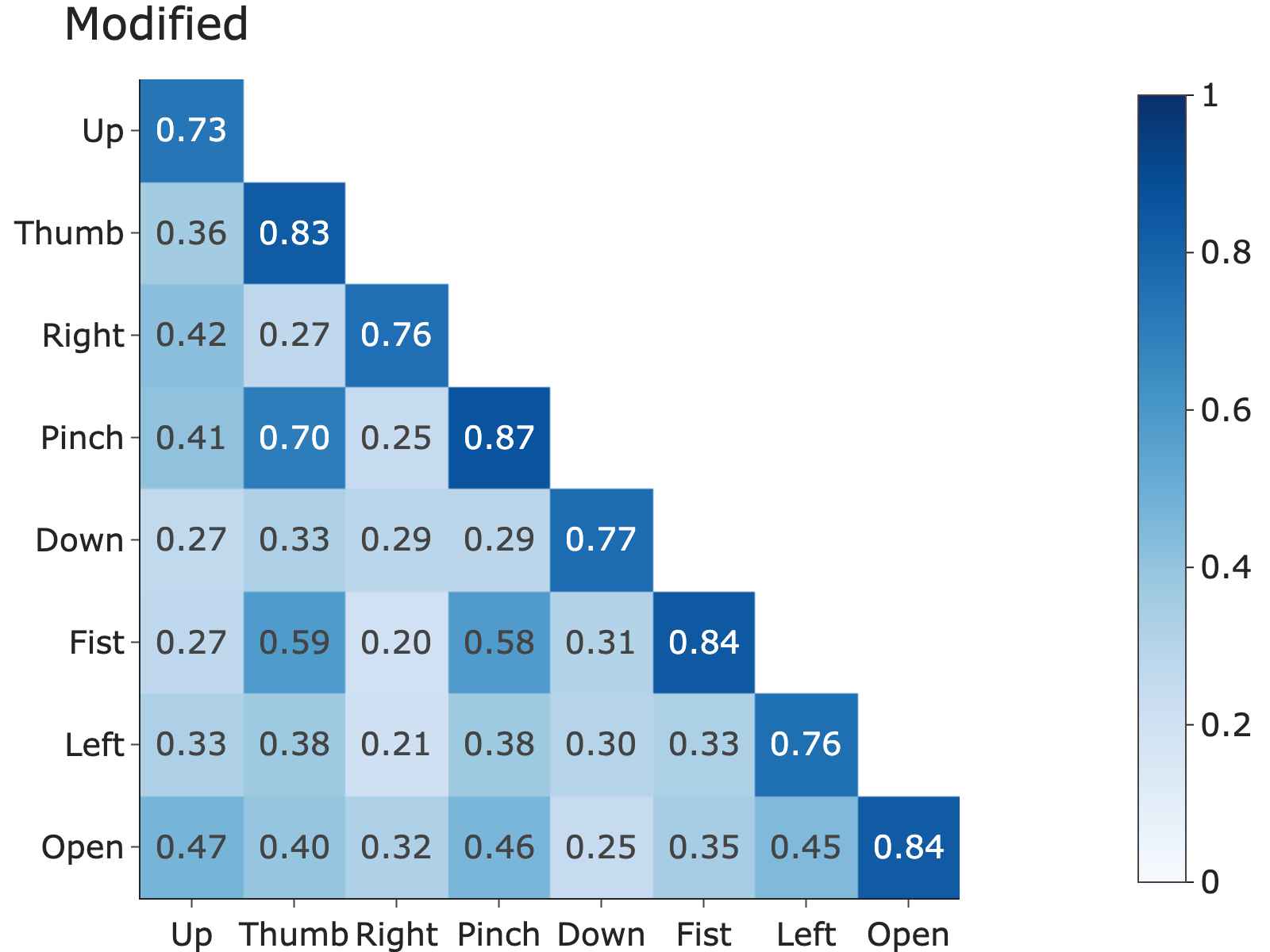}
    \hspace{-0.5mm}
    \includegraphics[width=0.04\linewidth]{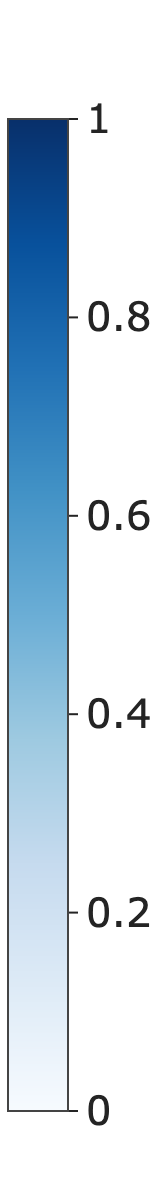}
    \includegraphics[width=0.30\linewidth,clip,trim={0 0 350 0}]{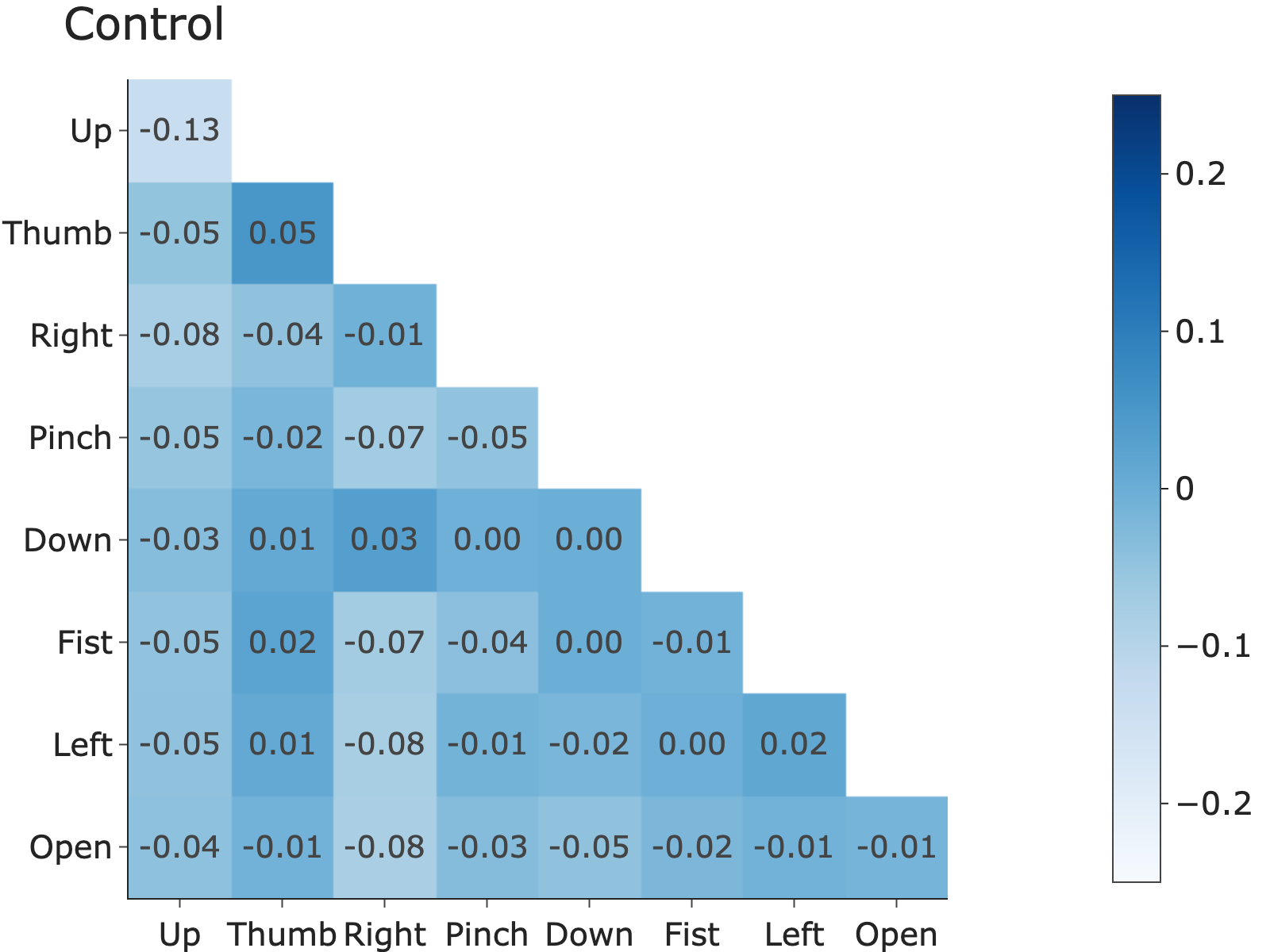}
    \includegraphics[width=0.30\linewidth,clip,trim={0 0 350 0}]{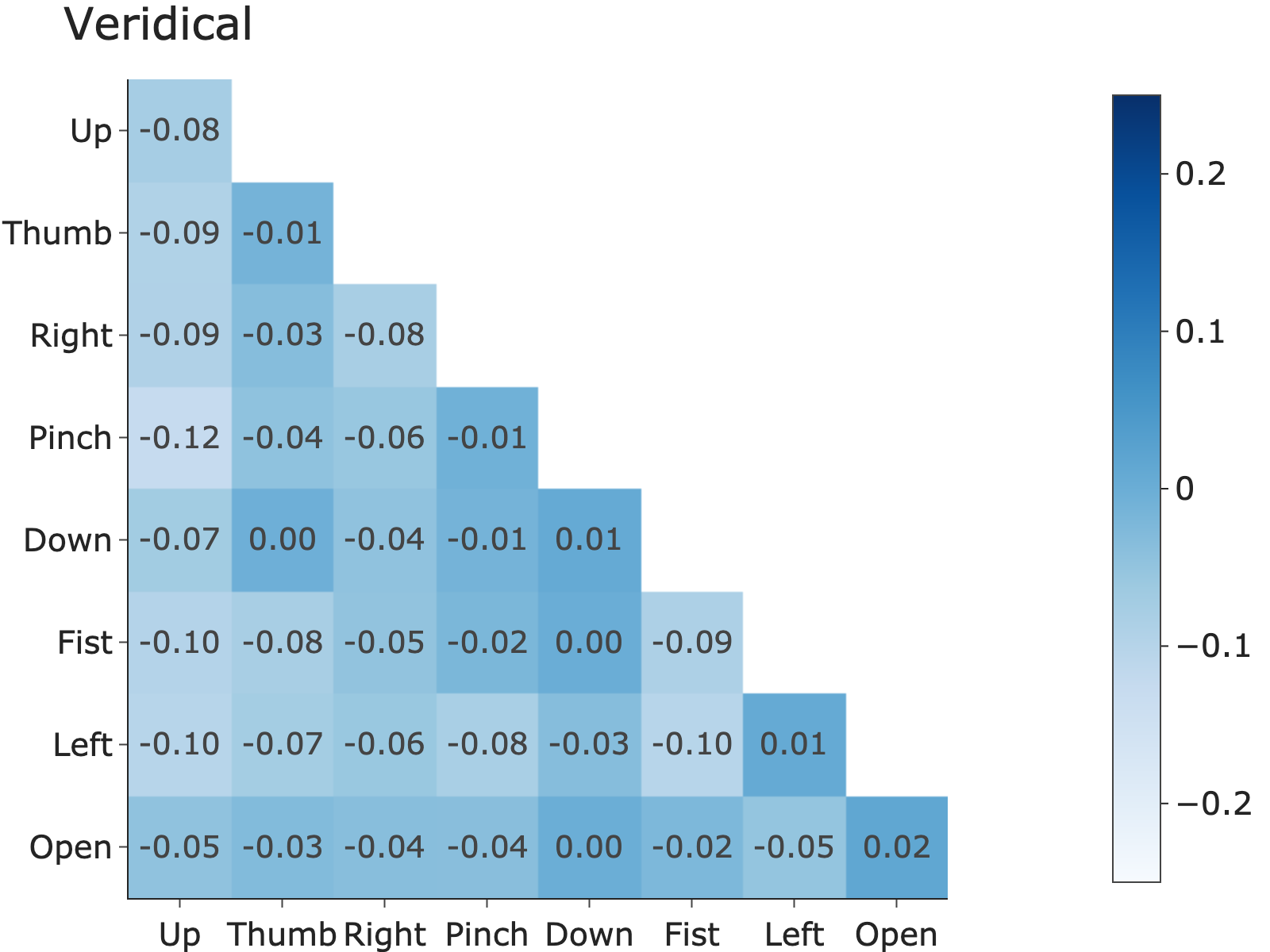}
    \includegraphics[width=0.30\linewidth,clip,trim={0 0 350 0}]{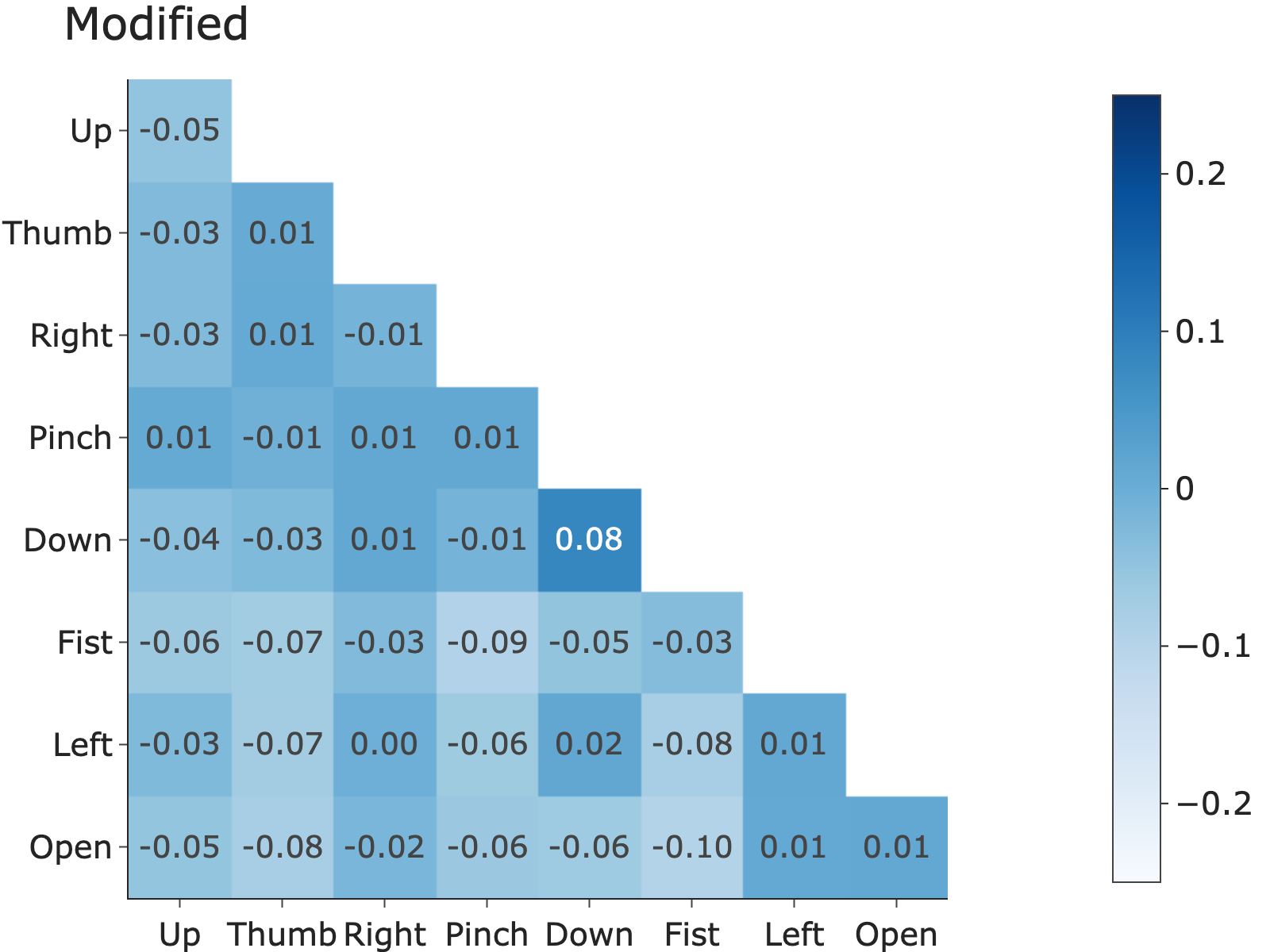}
    \includegraphics[width=0.05\linewidth]{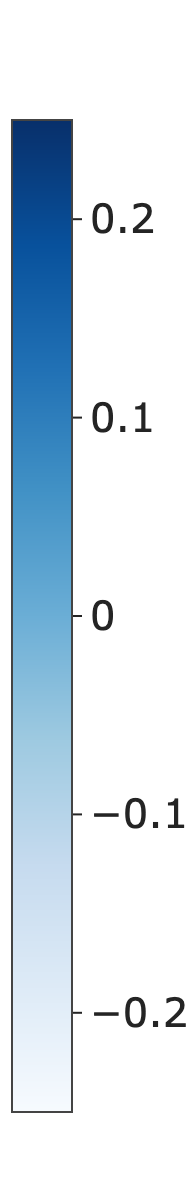}
    \caption{Normalized Class Similarity Matrices. Top row: Raw similarities from block four (free games, see section~\ref{sec:Block_Four_Free_Games}). Class similarity matrix $D$ is computed for each subject, normalized to $[0, 1]$, and then averaged across subjects in a group. Large values on the diagonal indicate tight clusters for each class. Small values off-diagonal indicate well-separated clusters. Bottom row: Change in similarity matrix from baseline $\Delta D$, as described in Equation~\ref{eqn:changes_from_baseline}. Positive values indicate pairs that became closer in feature space, compared to baseline; subjects whose structure improved would show positive values on the diagonal and negative values off-diagonal. See Section~\ref{sec:Feature_Space_Class_Structure} for further details. Left: Control group. Middle: Veridical feedback. Right: Modified feedback. Upper triangular parts are omitted due to symmetry.}
    \label{fig:similarity_matrices}
\end{figure*}
Figure~\ref{fig:similarity_matrices} shows the average normalized class similarity matrix of each group. 
By examining the diagonal entries, we can understand the repeatability of gestures (i.e. the similarity between items of the same class); by examining the off-diagonal entries, we can understand the separability of gestures (i.e. the similarity across different classes).
As described previously, a ``desirable" pattern for easy downstream classification (in which the subject produced consistent and well-separated gestures) would consist of larger entries on the diagonal and smaller entries off-diagonal. 

Each group demonstrated a consistent pattern in which the diagonal entries were brighter than the off-diagonal entries, indicating that the gestures were generally repeatable and well-separated. There was also a consistent pattern of bright off-diagonal cells, indicating high similarity between three specific gestures: ``pinch", ``fist", and ``thumb".
These patterns match well with the patterns visible in the class confusion matrices shown in Figure~\ref{fig:group_conf_mats}.
This correspondence between our similarity metrics and confusion matrices may indicate that our chosen similarity metric is well-suited to this setting and aligns well with model performance.

We did not observe any gross changes in the structure of class similarity between groups; note that such a change could have occurred if feedback affected gestures differently, and this effect may not have been visible by only inspecting the scalar $d_{\textsc{sep}}$ metric.

\section{Discussion and Future Work}\label{sec:Discussion}
This study tested the potential of modified continuous feedback of model performance in a gamified user interface for rapid user training on a sEMG-based gesture recognition system for controlling actions on a computer display. 

We hypothesized that we could use manipulation of feedback about the gesture class probabilities in a short (4-minute) online learning session to shape user behavior in a manner that would increase the separation between muscle activation patterns of different gestures and increase the accuracy of model performance on future attempts. Overall, our results demonstrate that a short user training session using modified feedback has the potential to increase post-calibration performance (accuracy and class separation relative) when compared to veridical feedback and a no-feedback control. 

\subsection{User Calibration} 
Despite the emergence of research into methods for co-adaptive learning for sEMG-based gesture recognition, there have been few investigations specifically testing the effect of user training as a means of rapid calibration. Numerous studies have shown that extended user training on an sEMG-based controller results in significant gains in performance \cite{he2015user, powell2013user, bunderson2012quantification}. The majority of these studies have found that increased model performance was accompanied by changes in muscle activation patterns that are theoretically favorable to better classification (such as improvements in class separability, variability, or repeatability). However, feature space characteristics of class distributions are not necessarily predictive of classifier performance, and this relationship is likely strongly dependent on the classifier used and the relationship between training and test data. For example, a recent investigation showed that the relationship between performance and feature-space metrics can be complex; these authors found that the real-time performance of an LDA classifier was only weakly correlated with class separability, but was not correlated with variability or repeatability \cite{franzke2020exploring}. Krasoulis et. al. first demonstrated that short-term adaptation through biofeedback user training could positively impact prosthetic finger control using sEMG-based decoding \cite{krasoulis2019effect}. Our results demonstrate that subjects who received modified live feedback experienced a significant increase in classification accuracy. We also found that both veridical and modified feedback provided a trend of improvement in our feature space metric $d_{\textsc{sep}}$, though this effect was not statistically significant.

\subsection{Influence of Feedback Manipulation on User Behavior.}
In our experiments, the Modified feedback group showed the largest change in classification accuracy and class separability. Flattening of the class probabilities as was done here can be considered a form of error augmentation, since subjects were led to believe that the separation between classes was smaller than it actually was. This approach is most closely related to techniques involving feedback with ``error amplification,'' which has been studied extensively.  Feedback of performance outcomes that are worse than actual performance (i.e. error amplification) has been found to expedite motor adaptations to novel task constraints compared to accurate feedback \cite{domingo2010effects,patton2006evaluation}. Amplification of task errors has also shown promise as an approach to facilitate motor recovery in patients with neurological disorders \cite{israely2016error,adamovich2009sensorimotor}. Faster or more complete learning with error amplification has been attributed to more brain processes associated with greater attention to execution of the motor task \cite{boussaoud1997primate,jueptner1998review,shirzad2012error} and reduction of sensorimotor noise \cite{hasson2016neuromotor}. We speculate that improvement in classification accuracy with Modified feedback in this study may be a product of similar mechanisms.

\subsection{Selected Gestures}
We selected gestures that mimicked the manipulation of commonplace items such as remote controls and cell phones. No subject commented that the gestures were unfamiliar or difficult to perform.
Directional gestures using wrist movements (``Up", ``Down", ``Left", ``Right") were generally more separable and yielded higher classification accuracy compared to gestures using grasping movements (``Pinch", ``Thumb", ``Open", ``Fist"). 
The extrinsic hand muscle groups used by each of these grasping gestures are similar, which may explain why subjects had a difficult time performing them accurately while also creating separation in muscle activation patterns.
Thus the feature-space similarity that we observed for these grasping gestures is somewhat expected. 

\subsection{Limitations}
There were several limitations of the current work that may have affected the results and interpretations. Only a single classification model was used. Several machine learning methods, including artificial neural networks, linear discriminant analysis, support vector machines (SVM), and Gaussian mixture models have been previously used for sEMG-based control. The choice to use a model based on SVM and logistic regression was due to its simplicity and the popularity of SVM for this application. It is possible that the choice of classifier model affects both calibration accuracy and the way that users explore the mapping of muscle activation to gestures. Nevertheless, the user training scheme employed here likely has general benefits for use and understanding of human co-adaptive behavior. 

There are a number of possible changes in the signal processing pipeline that may yield improvements in overall model performance. The active window for feature extraction may be tuned, and additional features such as time-frequency domain or higher-dimensional feature vectors may be extracted. The selected features (RMS, and median frequency) were chosen based on their common use for sEMG-based gesture classification and initial pilot testing. Future work should evaluate how sEMG feature selection affects user training. 

\subsection{Designing Improved Feedback}
Only a single type of feedback manipulation was tested. We used a feedback manipulation that flattened probabilities across classes, making it more difficult to achieve a correct classification. This approach was selected as it was expected that participants would respond by increasing the separation between muscle activation patterns for different gestures. While we observed a non-significant trend of improvement in class separation, the manipulation was not directly optimized for this purpose. Future research should explore the optimization of feedback manipulation for shaping user behavior during co-adaptive sEMG-gesture recognition. Adaptive feedback manipulation based on user and model performance characteristics to target specific class confusions is an attractive future direction. Further improvement may come from iterating between rounds of visual feedback to induce human learning, and rounds of model re-training using the subject's most recent data.
The approach we used was a form of modified knowledge of results; future work could explore using modified knowledge of performance by giving the user feedback about feature space characteristics such as distance between the current feature vector and a representative item from the target class, or aggregate feature metrics describing properties like separability and repeatability.

\bibliography{references.bib}
\end{document}